\documentclass[useAMS,usenatbib]{mnras}
\usepackage{journals}
\usepackage{graphicx}
\usepackage{color}
\usepackage{times}
\usepackage{hyperref}
\usepackage{amsmath}

\title[The impact of DWs on the distribution of SNe]
{The impact of spiral density waves on the distribution of Supernovae}
\author[A.~G.~Karapetyan~et~al.]{A.~G.~Karapetyan,$^{1}$\thanks{E-mail:
karapetyan@bao.sci.am (AGK); hakobyan@bao.sci.am (AAH)}
A.~A.~Hakobyan,$^{1}$\footnotemark[1]
L.~V.~Barkhudaryan,$^{1}$
\newauthor
G.~A.~Mamon,$^{2}$
D.~Kunth,$^{2}$
V.~Adibekyan$^{3}$
and M.~Turatto$^{4}$
\\
$^{1}$Byurakan Astrophysical Observatory, 0213 Byurakan, Aragatsotn province, Armenia\\
$^{2}$Institut d'Astrophysique de Paris, Sorbonne Universit\'{e}s, UPMC Univ Paris 6 et CNRS, UMR 7095, 98 bis bd Arago, F-75014 Paris, France\\
$^{3}$Instituto de Astrof\'{i}sica e Ci\^{e}ncia do Espa\c{c}o, Universidade do Porto, CAUP, Rua das Estrelas, P-4150-762 Porto, Portugal\\
$^{4}$INAF -- Osservatorio Astronomico di Padova, Vicolo dell'Osservatorio 5, I-35122 Padova, Italy}
\begin{document}
\date{Accepted ---. Received ---; in original form ---}

\pagerange{\pageref{firstpage}--\pageref{lastpage}} \pubyear{2018}

\maketitle

\label{firstpage}

\begin{abstract}
  We present an analysis of the impact of spiral density waves (DWs) on the radial and
  surface density distributions of supernovae (SNe) in host galaxies
  with different arm classes.
  We use a well-defined sample of 269 relatively nearby,
  low-inclination, morphologically non-disturbed and unbarred Sa--Sc galaxies
  from the Sloan Digital Sky Survey, hosting 333 SNe.
  Only for core-collapse (CC) SNe, a significant difference appears when comparing their
  $R_{25}$-normalized radial distributions in long-armed grand-design (LGD) versus non-GD (NGD) hosts,
  with that in LGD galaxies being marginally inconsistent with an exponential
  profile, while SNe~Ia exhibit exponential surface density profiles
  regardless of the arm class.
  Using a smaller sample of LGD galaxies with estimated corotation radii ($R_{\rm C}$),
  we show that the $R_{\rm C}$-normalized surface density distribution of
  CC~SNe indicates a dip at corotation. Although not statistically significant,
  the high CC~SNe surface density just inside and outside corotation
  may be the sign of triggered massive star formation by the DWs.
  Our results may, if confirmed with larger samples,
  support the large-scale shock scenario induced by spiral DWs in LGD galaxies,
  which predicts a higher star formation efficiency around the shock fronts,
  avoiding the corotation region.
\end{abstract}


\section{Introduction}
\label{intro}

The spiral arm structure of star-forming disc galaxies
was explained in the framework of density wave (DW) theory
by the pioneering work of \citet[][]{1964ApJ...140..646L}.
According to this theory, semi-permanent spiral patterns
especially in grand-design (GD) galaxies, i.e. spiral galaxies with prominent and well-defined spiral arms,
are created by long-lived quasi-stationary DWs.
Despite an excellent progress of the theory
\citep[for recent comprehensive reviews, see][]{2014PASA...31...35D,2016ARA&A..54..667S},
there are many disputes on the lifetime of spiral patterns, and
the ability of DWs to generate large-scale shocks and trigger star formation,
as originally proposed by \citet[][]{1969ApJ...158..123R}.
For example, the simulations by \citet[][]{2011MNRAS.410.1637S} manifest short-lived patterns.
In another example, using a multiband analysis for some GD galaxies,
\citet{2010ApJ...725..534F} found that there is no shock trigger,
and that the spiral arms just reorganize the material from the disc out of which stars form
\citep[see also][]{2012A&A...542A..39G}.

Nevertheless, the results of many other studies are consistent with the picture
where the DWs cause massive star formation to occur by compressing gas clouds
as they pass through the spiral arms of GD galaxies
\citep[e.g.][]{1990ApJ...349..497C,1996A&A...308...27K,2002MNRAS.337.1113S,2009A&A...499L..21G,
2009ApJ...694..512M}.
For example, using H$\alpha$ direct imaging accompanied with broad-band images in $R$ and $I$ bands,
\citet{2013AA...560A..59C} studied the distribution of H~{\footnotesize II} regions of spiral arms and
found clear evidence for the triggering of star formation in
the sense of a high density of H~{\footnotesize II} regions at the fixed radial ranges in some GD galaxies.
Recently, \citet{2016ApJ...827L...2P} showed that pitch angle of galaxies is statistically more tightly wound,
i.e. smaller, when viewed in the light from the evolved/older stellar populations.
Both the results, complementing each other, are in excellent agreement with the prediction of theory
that stars are not only born in the DW but also move out of it as they age
\citep[see also most recent results by][]{2018MNRAS.478.3590S}.

An alternative to the DW theory is the idea of reorganization of
the distribution of H~{\footnotesize II} regions in multiple arms of differentially rotating disc
with star formation processes generated by the stochastic self-propagating method developed by
\citet[][]{1976ApJ...210..670M} and \citet[][]{1978ApJ...223..129G}.
This mechanism is supposed to work in non-GD (NGD) galaxies, producing flocculent spiral arms.

In the context of above-mentioned scenarios,
the main goal of this article is to study the possible impact of spiral DWs (triggering effect)
on the distribution of supernovae (SNe) in discs of host galaxies,
when viewing in the light of different nature of Type Ia and core-collapse (CC) SNe progenitors.
Recall that Type Ia SNe result from stars with masses lower than $\sim7.5~M_{\odot}$
\citep[ages from $\sim0.5$~Gyr up to $\sim10$~Gyr, see][]{2012PASA...29..447M} in close binary systems,
while the progenitors of Types Ibc and II SNe,\footnote{{\footnotesize Traditionally, SNe of Types Ib and Ic,
including uncertain spectroscopic Type Ib/c, are denoted as SNe Ibc.
All these and other subtypes of CC~SNe, i.e. Ibc, II, IIb (transitional objects with observed properties close to SNe II and Ib),
and IIn (dominated by emission lines with narrow components) SNe,
arise from young massive stars with possible differences in their masses, metallicities, and ages
\citep[see e.g.][for more details]{2011MNRAS.412.1522S}.}}
collectively called CC~SNe, are massive
\citep[$M\gtrsim7.5~M_{\odot}$, see e.g.][]{2018ApJ...860...39W} young short-lived stars
\citep[from a few up to $\sim100$~Myr, see][]{2015PASA...32...19A,2018MNRAS.476.2629M,2018arXiv180501213X}.

The first attempt to study the distribution of SNe within the framework of DW theory
was performed by \citet{1973PASP...85..564M}.
Using the locations of 19 SNe, he suggested that stars in a spiral galaxy
are formed in a shock front on the inner edge of a spiral arm,
then drift across the arm as they age, predicting for SN progenitors (more likely for SNe II)
a short lifetime (a few million years) and high masses (a few tens of solar masses).
However, using the fractions of GD and flocculent galaxies
in a sample of 111 hosts with 144 SNe,
\citet{1986ApJ...311..548M} suggested that DWs do not greatly enhance
the massive star formation rate per unit luminosity of a galaxy,
mentioning that star formation in most galaxies may be dominated by stochastic processes.
Results similar to those in \citet{1973PASP...85..564M} were obtained also by
\citet{1996ApJ...473..707M} and \citet{2007AstL...33..715M} for Types II and Ibc SNe, respectively.
In other studies, different authors \citep[e.g.][]{1976ApJ...204..519M,1994PASP..106.1276B,2005AJ....129.1369P}
investigated the distribution of SNe relative to spiral arms of galaxies.
Such studies did not interpret their results within the DW theory
nor did they distinguish among various spiral arm classes \citep[ACs;][]{1987ApJ...314....3E} of SNe host galaxies.

Indeed, in our recent paper \citep{2016MNRAS.459.3130A},
we already studied the distribution of SNe relative to the spiral arms of their GD and NGD
host galaxies, using the Sloan Digital Sky Survey (SDSS) images from the $g$, $r$, and $i$ bands.
We found that the distribution of CC~SNe (i.e. tracers of recent star formation)
is affected by the spiral DWs in their host GD galaxies,
being distributed closer to the corresponding edges of spiral arms where large-scale shocks,
thus star formation triggering, are expected (see also farther in the text of Section~\ref{DWTInt}).
Such an effect was not observed for Type Ia SNe (less-massive and longer-lived progenitors) in GD galaxies,
as well as for both types of SNe in NGD hosts.
In this paper, we expand our previous work, and for the first time study the differences between
the radial distributions of SNe in unbarred Sa--Sc host galaxies with various spiral ACs.
In parallel, to check the triggering effect at different galactocentric radii,
we study the consistency of the surface density distribution of SNe
(normalized to the optical radii, and for a smaller sample also to corotation radii of hosts)
with an exponential profile in GD and NGD galaxies.

The layout of this article is the following.
In Section~\ref{sample}, we present sample selection and reduction,
and determination approach of spiral ACs.
The results and their interpretation within the framework of DW theory
are presented in Sections~\ref{resdiscus} and \ref{DWTInt}, respectively.
Section~\ref{concl} summarizes our conclusions.
To conform the values used in databases of our recent papers
\citep[][]{2012A&A...544A..81H,2014MNRAS.444.2428H,2016MNRAS.456.2848H,2016MNRAS.459.3130A},
a cosmological model with $\Omega_{\rm m}=0.27$, $\Omega_{\rm \Lambda}=0.73$,
and $H_0=73 \,\rm km \,s^{-1} \,Mpc^{-1}$ Hubble constant \citep{2007ApJS..170..377S}
are adopted in this article.

\section{The sample}
\label{sample}

In order to obtain a homogeneous dataset of structural features of SNe host galaxies,
including morphology, identification of bars and spiral ACs,
we compile the sample of this study in the same way as in \citet[][]{2012A&A...544A..81H},
being restricted to relatively nearby SNe with distances ${\leq {\rm 150~Mpc}}$.
The whole compilation, reduction, and classification procedures are given below.

\subsection{Sample selection and reduction}
\label{sample1}

We used the updated version of the Asiago Supernovae Catalogue \citep[ASC;][]{1999A&AS..139..531B},
which at the time of writing the article includes SNe detected before 1~July 2017.
To identify SNe host galaxies, we cross-matched the coordinates of
all classified Type Ia and CC (Ibc and II) SNe from the ASC
with the footprint of the SDSS Data Release 13 \citep[DR13;][]{2017ApJS..233...25A}.
We then classified the identified host galaxies according to \citet[][]{2012A&A...544A..81H}
and selected only Sa--Sc types,\footnote{{\footnotesize Many of the identified host galaxies
are already listed in database of \citet[][]{2012A&A...544A..81H},
which is based on the SDSS DR8. Here, because we added new SNe, for homogeneity we redid
the entire reduction for this restricted sample based only on DR13.}}
since it is known that both GD and NGD shapes are well represented
in Sa--Sc spirals \citep[e.g.][]{2013JKAS...46..141A,2017MNRAS.471.1070B}.

We excluded all barred galaxies from our sample to eliminate
the effect of substantial suppression of massive star formation
in the radial range swept by strong bars \citep[e.g.][]{2009A&A...501..207J,2018MNRAS.474.3101J},
i.e. the observed suppression of CC~SN numbers inside the bar radius \citep[][]{2016MNRAS.456.2848H},
and study only the expected impact of the DWs on the distribution of
SNe.\footnote{{\footnotesize It is important to note that in some SN host galaxies
we may not detect tiny bars with
lengths shorter than a tenth of the optical disc
\citep[][]{2014MNRAS.444.2428H}.
However, by the inner truncation of host discs (see Subsection~\ref{resdiscus_sub1})
we exclude any possible impact of these bars on the distribution of SNe.}}
In addition, we removed host galaxies with strong
morphological disturbances according to \citet[][]{2014MNRAS.444.2428H},
i.e. interacting, merging and post-merging/remnant cases,
which may add significant distortion into the SN distribution in discs of galaxies.

For the remaining SNe host galaxies, the next step is the measurement of their photometry and geometry.
Following \citet[][]{2012A&A...544A..81H}, we constructed $25~{\rm mag~arcsec^{-2}}$ isophotes
in the SDSS DR13 $g$-band, and then visually fit onto each isophote an elliptical aperture.
We then measured the major axes ($D_{25}$), elongations ($a/b$), and position angles (PA) of galaxies.
In our analysis, we used the $D_{25}$ corrected for Galactic \citep[][]{1998ApJ...500..525S} and
host galaxy internal extinction \citep[][]{1995A&A...296...64B}.
Finally, we calculated the inclinations of host galaxies using elongations and
morphological types, following the method presented in \citet[][]{1997A&AS..124..109P}.
These procedures are explained in detail in \citet[][]{2012A&A...544A..81H}.

We also removed highly inclined galaxies ($i > 60^\circ$),
because at these inclinations strong absorption and projection effects play a destructive role
in discovering SNe \citep[e.g.][]{1997ASIC..486...77C} and
correcting their radial distribution for inclination of host disc \citep[e.g.][]{2016MNRAS.456.2848H}.
Moreover, it is difficult to classify highly inclined galaxies
and determine their barred structure \citep[see review by][]{2013pss6.book....1B}.

After these operations, we obtained
353 SNe within 285 host galaxies with the aforementioned restrictions.

\subsection{Determination of spiral arm classes}
\label{sample2}

\begin{figure*}
\begin{center}$
\begin{array}{@{\hspace{0mm}}c@{\hspace{2mm}}c@{\hspace{2mm}}c@{\hspace{2mm}}c@{\hspace{0mm}}}
\includegraphics[width=0.235\hsize]{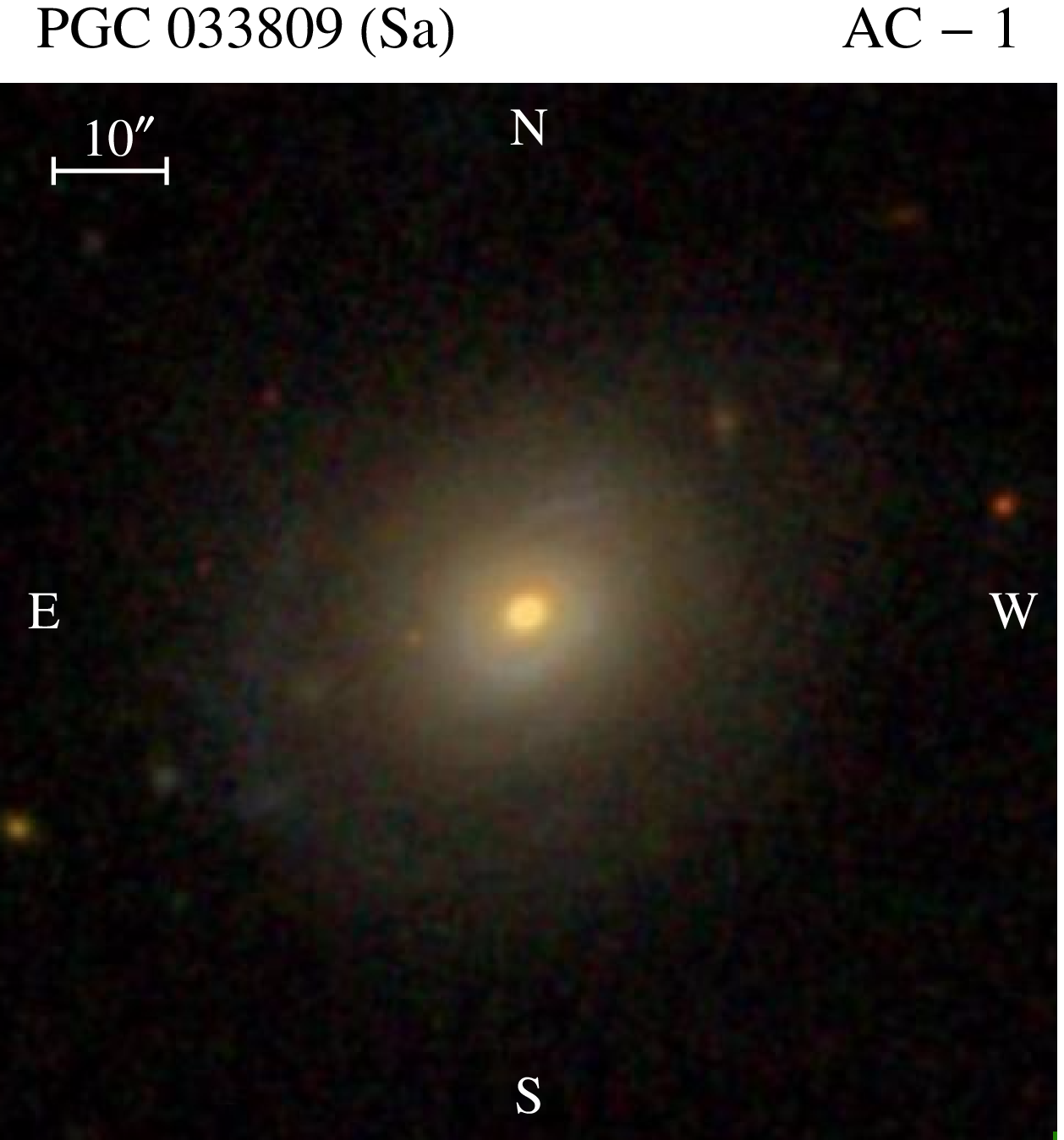} &
\includegraphics[width=0.235\hsize]{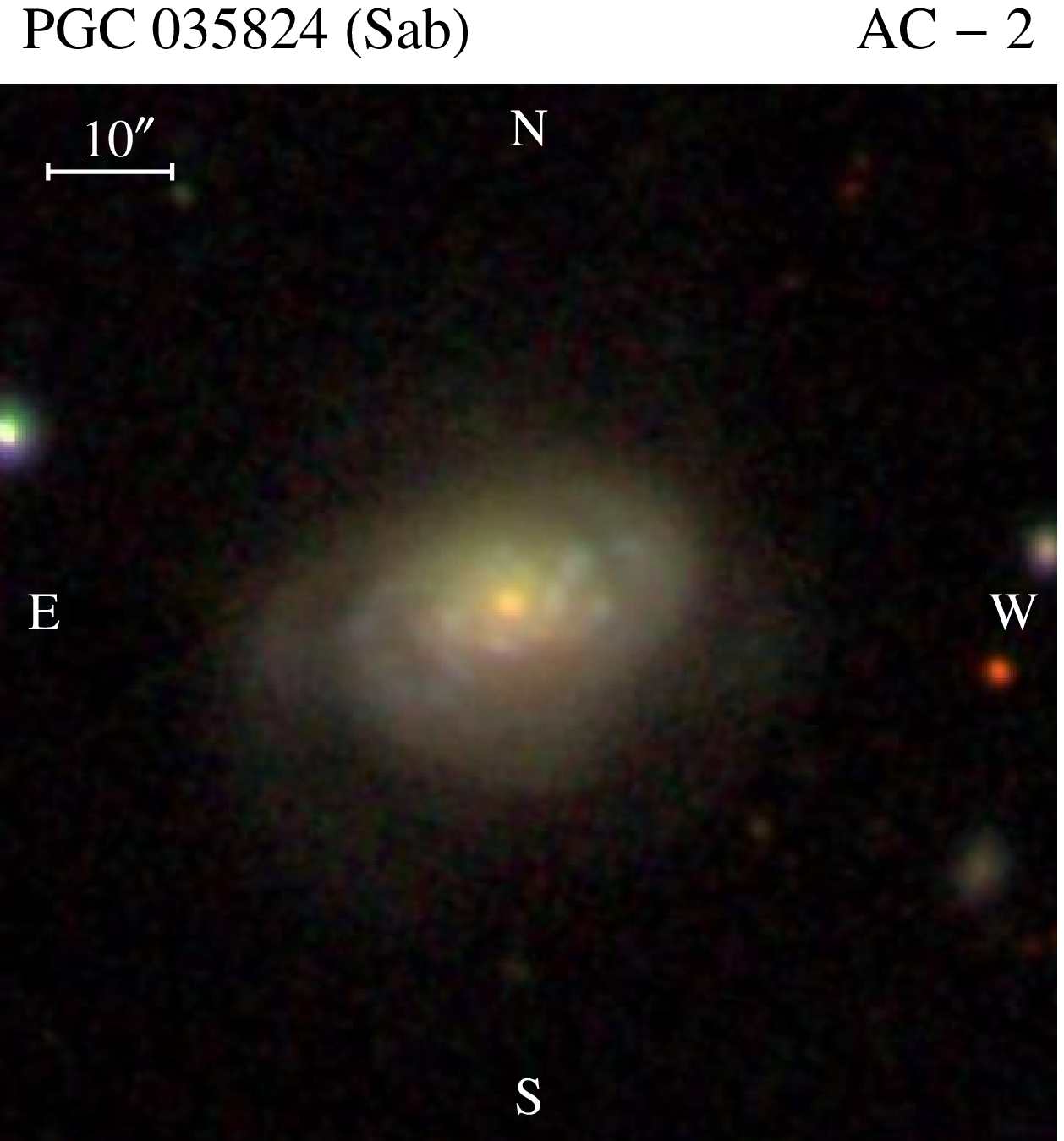} &
\includegraphics[width=0.235\hsize]{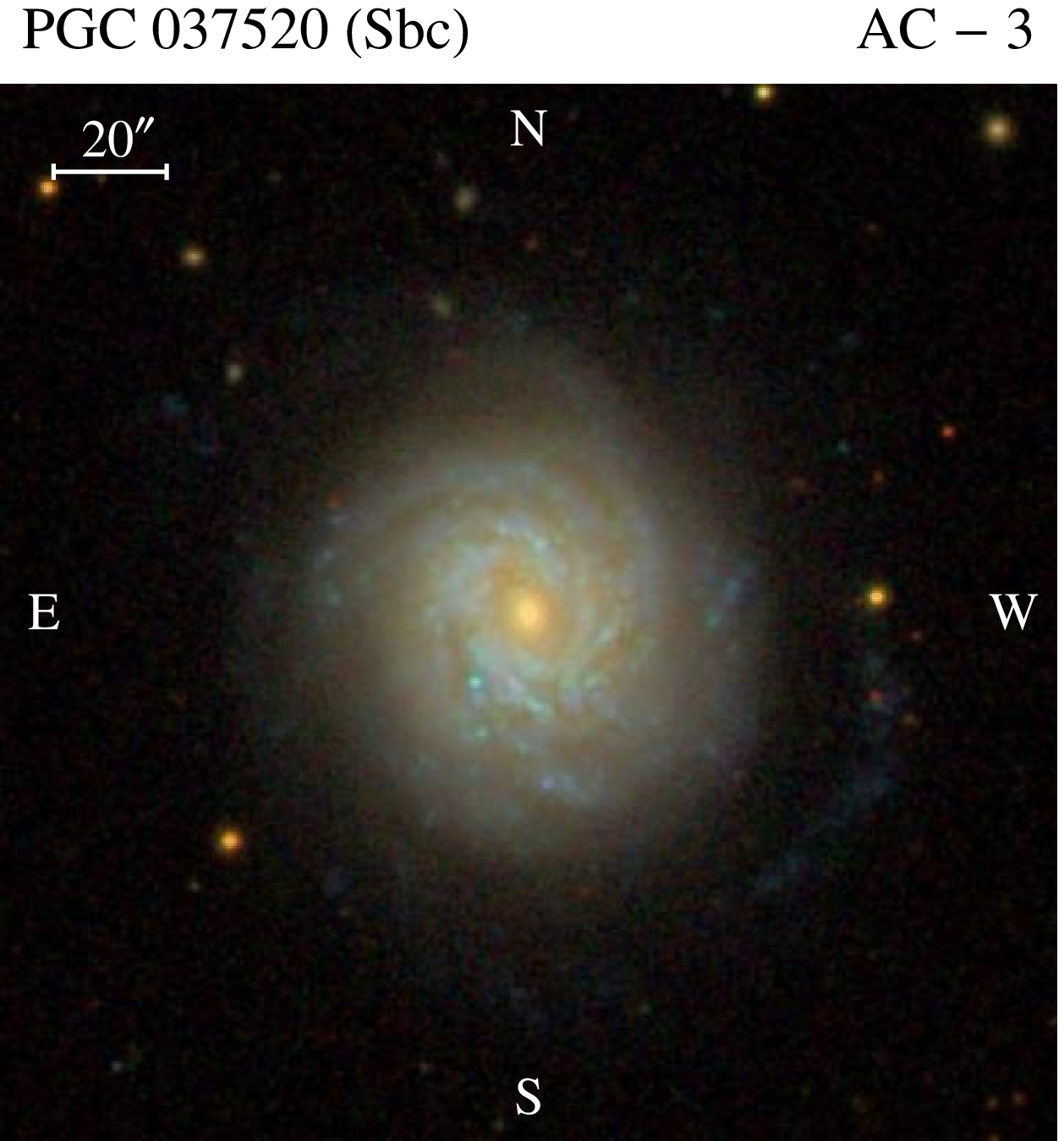} &
\includegraphics[width=0.235\hsize]{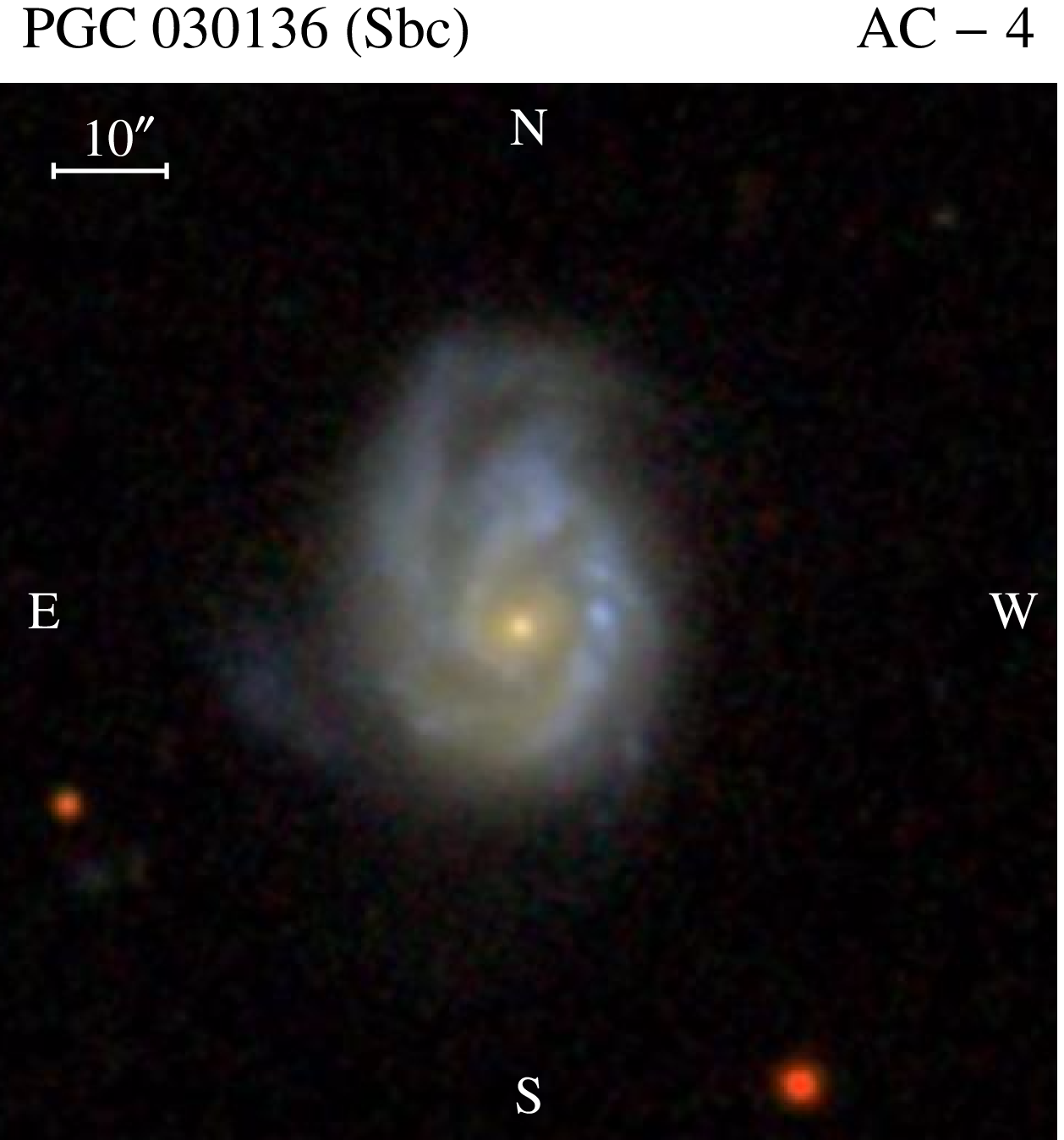} \\
\includegraphics[width=0.235\hsize]{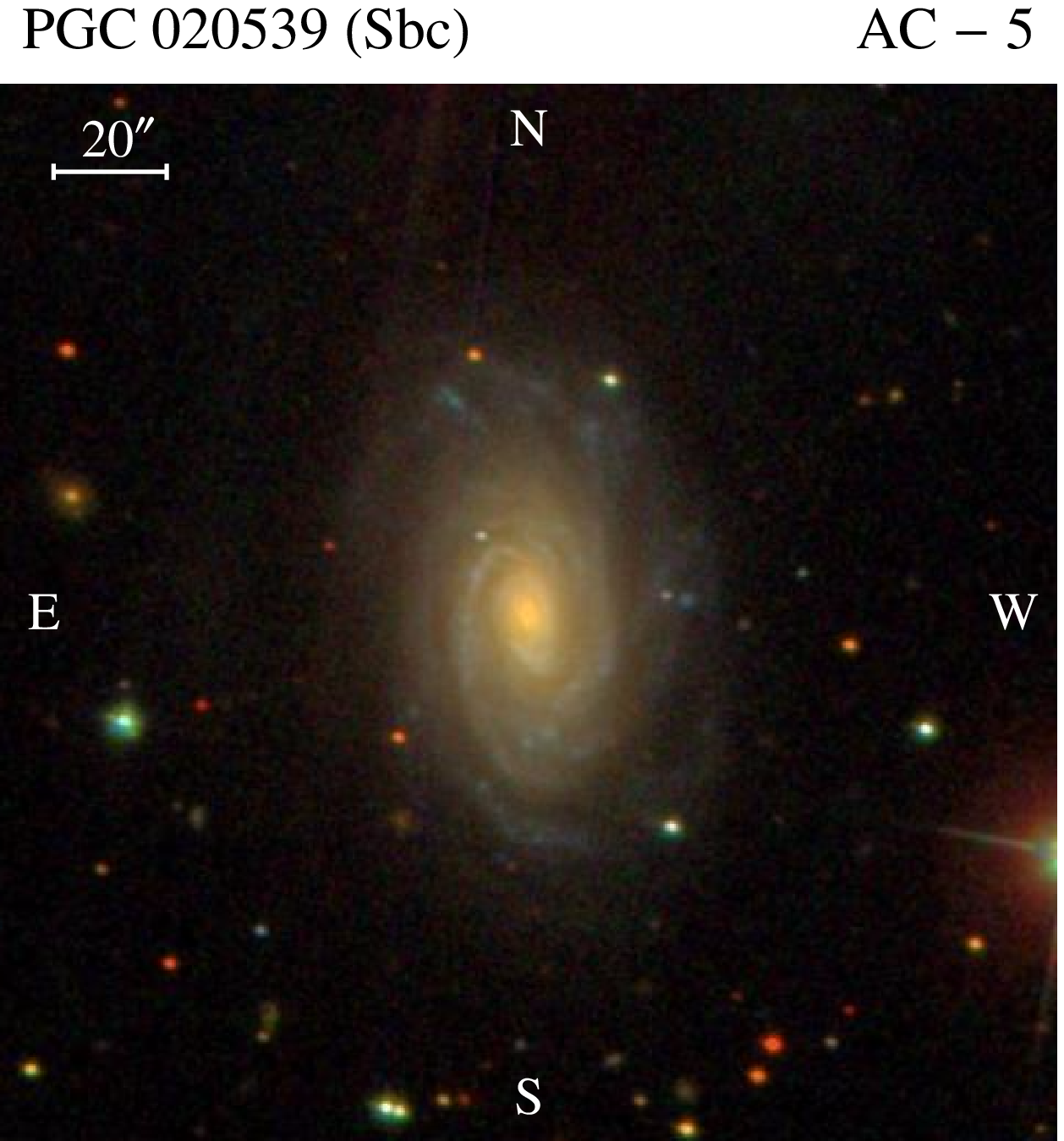} &
\includegraphics[width=0.235\hsize]{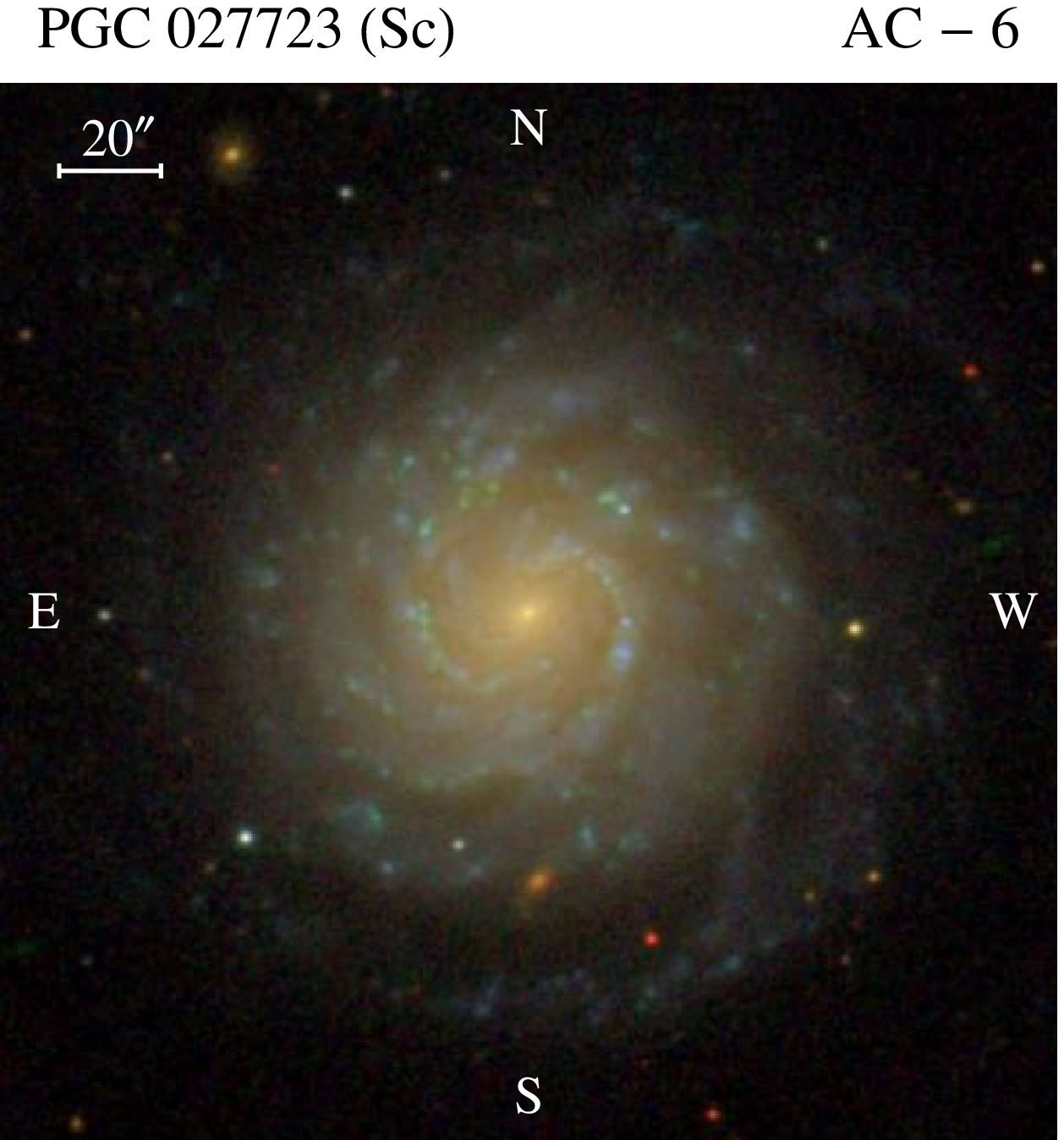} &
\includegraphics[width=0.235\hsize]{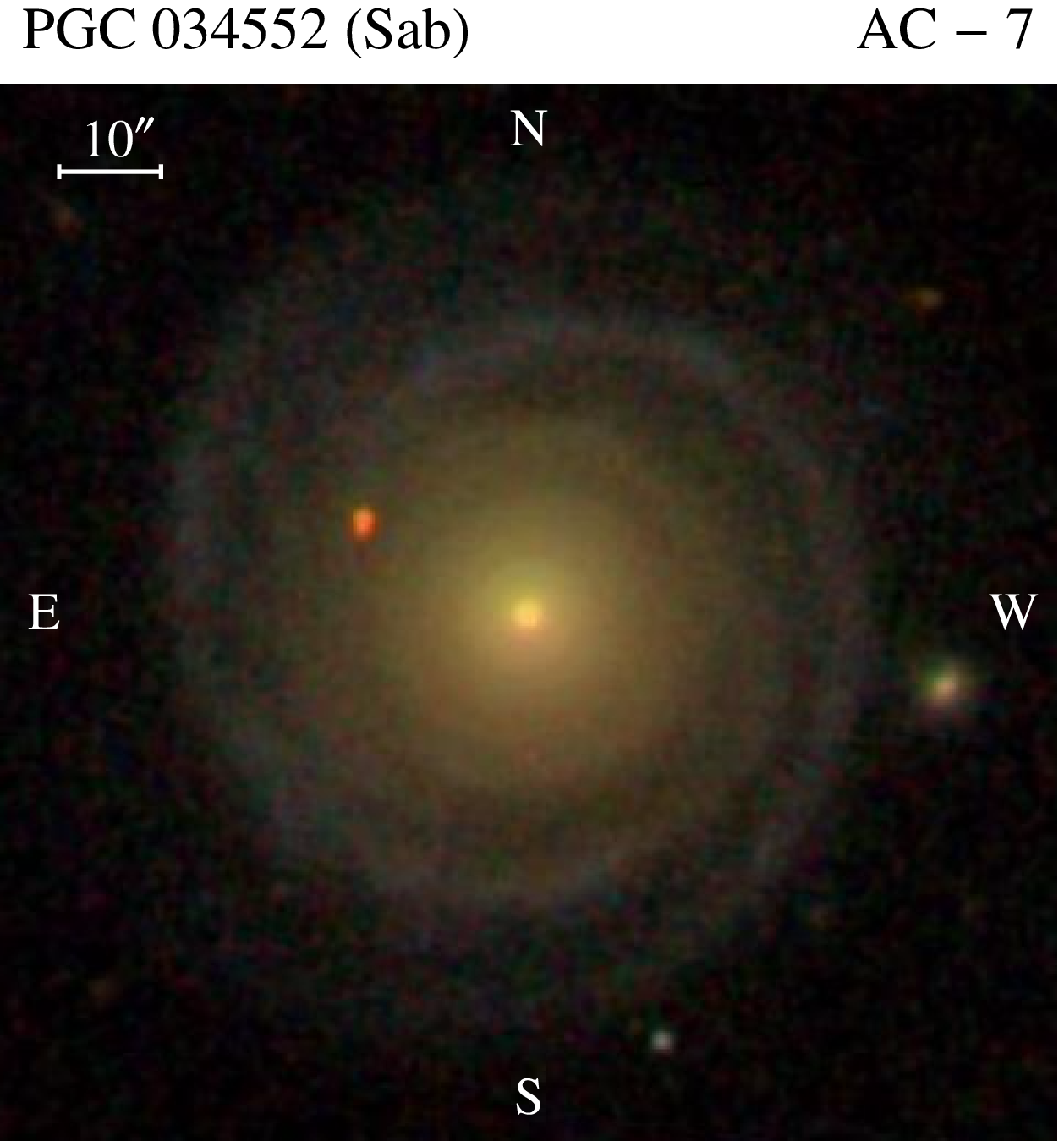} &
\includegraphics[width=0.235\hsize]{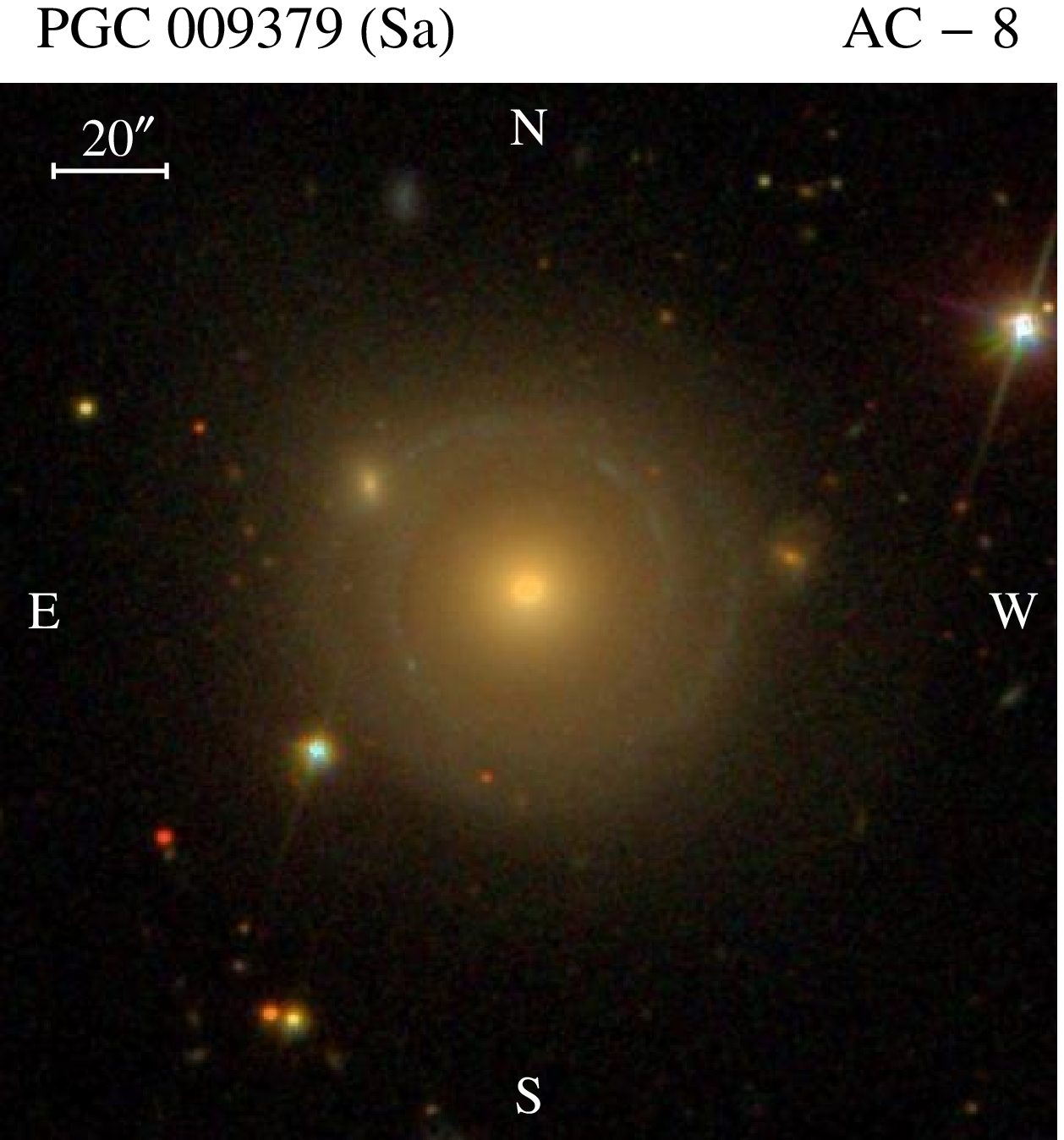} \\
& \includegraphics[width=0.235\hsize]{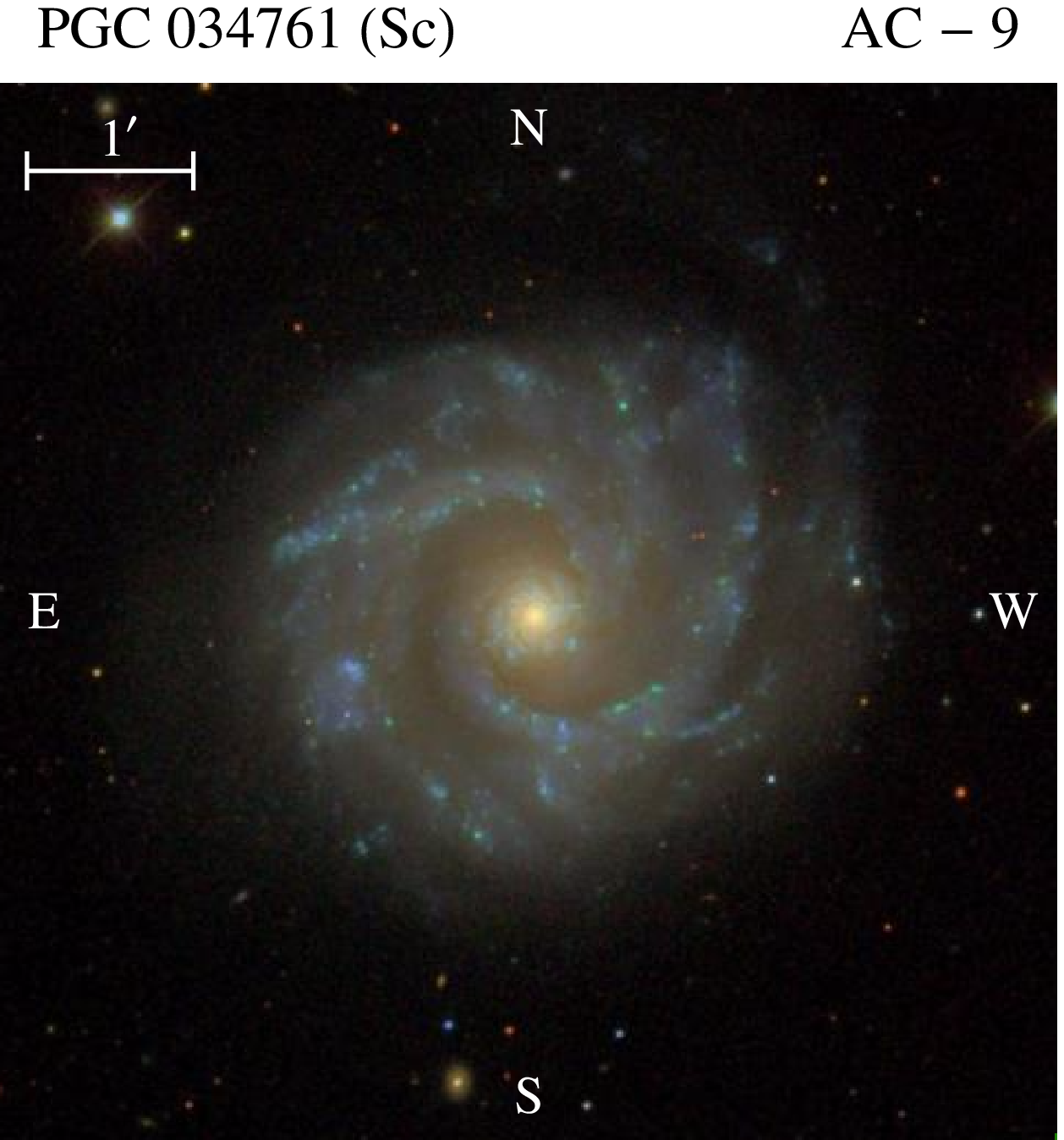} &
\includegraphics[width=0.235\hsize]{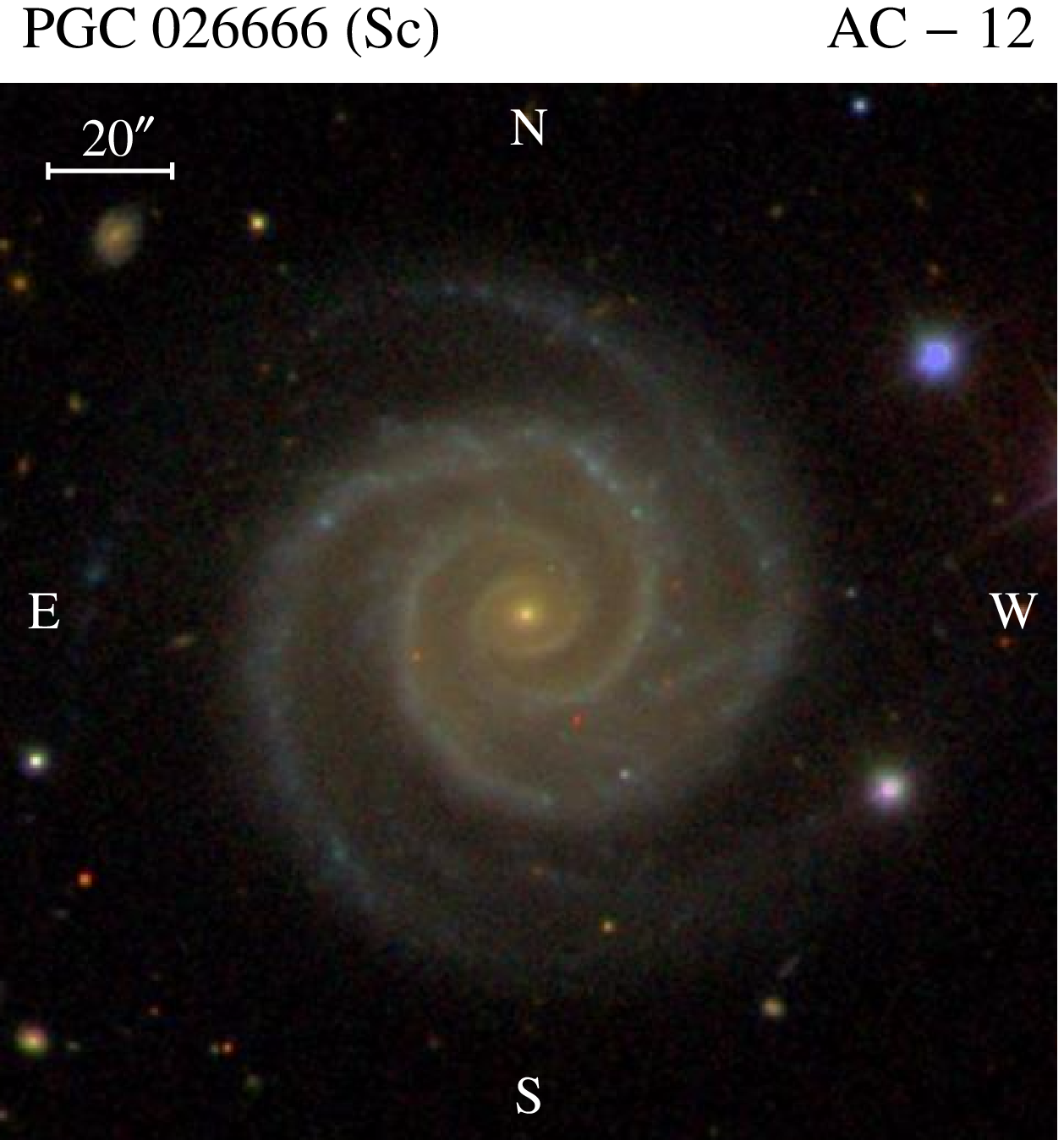} &
\end{array}$
\end{center}
\caption{SDSS images representing examples of unbarred Sa--Sc host galaxies
         with different arm classes (ACs) according to \citet[][]{1987ApJ...314....3E}.
         The Principal Galaxy Catalogue (PGC) objects' identifiers,
         morphological types (in parentheses), and ACs are listed at the top.
         In all images, north is up and east is to the left.}
\label{ArmClassexamples}
\end{figure*}

Following our recent study \citep[][]{2016MNRAS.459.3130A},
we determined ACs of 285 host galaxies (unbarred Sa--Sc types) with $i \leq 60^\circ$
according to the classification scheme by \citet[][]{1987ApJ...314....3E}.
To accomplish this, we used the background subtracted and photometrically calibrated
$g$-band\footnote{{\footnotesize Among the SDSS $g$, $r$, and $i$ bands with
good signal-to-noise ratio, the arm-interarm contrast is the highest in the $g$-band,
as it traces the young stellar populations in the spiral arms \citep[see][]{2016MNRAS.459.3130A}.}}
SDSS images, as well as the RGB colour images from
the $g$, $r$, and $i$ SDSS data channels.
We assigned ACs
according to the flocculence, regularity, and shapes of the spiral arms.
The SDSS three-colour images representing examples of SN host galaxies
with different ACs can be found in Fig.~\ref{ArmClassexamples}.
Below, we describe these classes in detail according to \citet[][]{1987ApJ...314....3E}.

Galaxies with AC 12 contain two long symmetric arms, and
the ones with AC 9 have two symmetric inner arms, multiple long and continuous outer arms.
The underlying mechanism that explains the lengths of arms and their global symmetry in
these galaxies is most probably a DW, dominating the entire optical disc \citep[e.g.][]{1992ApJS...79...37E}.
We denote galaxies with ACs 9 and 12 as long-armed GD (LGD) galaxies.

Galaxies with AC 5 have two symmetric short arms in the inner region and irregular outer arms.
The AC 6 is like AC 5 in the inner disc region, however with feathery ringlike outer structure.
The short inner symmetric arms in these galaxies might be explained by the DW mechanism,
dominating only in the inner part of the optical disc \citep[e.g.][]{1992ApJS...79...37E}.
We denote galaxies with ACs 5 and 6 as short-armed GD (SGD) galaxies.

Galaxies with AC 1 are described by chaotic, fragmented and unsymmetric arms,
AC 2 is fragmented spirals arm pieces with no regular pattern,
AC 3 is fragmented arms uniformly distributed around the galactic centre.
Galaxies with AC 4 have only one permanent arm, otherwise fragmented arms.
All these flocculent galaxies (ACs 1-4) appear to lack global DWs,
instead their spirals may be sheared self-propagating star formation regions
\citep[see review by][and references therein]{2013pss6.book....1B}.
We denote galaxies with ACs 1-4 as NGD galaxies.

Galaxies with AC 7 have two symmetric long outer arms, feathery or irregular inner arms.
In these galaxies, the DWs play a role, most probably, only in
the outer part of the optical disc
\citep[see review by][and references therein]{2013pss6.book....1B}.
In our study, due to the small number statistics (especially for CC~SNe),
these galaxies are not denoted to a separate class.
We have only 11 Type Ia and 6 CC~SNe in these hosts.
On the other hand, because of the different placement of DWs,
it is inadvisable to mix them with other classes.
Therefore, we simply omit them from the sample.

Galaxies with AC 8 have tightly wrapped ringlike arms.
These ringlike arms (rings and pseudorings) are thought to be related to the gathering of material
near dynamical resonances in the disc \citep[see review by][]{2013pss6.book....1B}.
Because of the different structural feature and small number statistics (only 3 Type Ia SNe),
we omit these galaxies from the sample as well.

Finally, according to \citet[][]{1987ApJ...314....3E}, ACs 10 and 11 were previously reported to be
barred galaxies and objects with close neighbors, respectively, and are no longer used.

In the present study, we mainly used these broad classes:
LGD (AC 9, 12), SGD (AC 5, 6), and NGD (AC 1-4).
Table~\ref{table_SN_morph} presents the distributions of 333 SN types among various
morphological types of the broad ACs of host galaxies.
The number of individual host galaxies is 269.
The mean distance of the galaxies is 82~Mpc
(standard deviation is 39~Mpc).
The mean $D_{25}$ of the hosts is 120 arcsec with the minimum value of 23 arcsec.
In Table~\ref{table_SN_morph}, we present the numbers of Types Ibc and II SNe separately.
However, to increase statistical significance of our results
(especially in Section~\ref{DWTInt}), we combined SNe~Ibc and II into a single CC~SNe class.

\begin{table}
  \centering
  \begin{minipage}{61mm}
  \caption{Numbers of SNe at distances $\leq 150~{\rm Mpc}$ in unbarred Sa--Sc
           hosts with inclinations $i\leq 60^\circ$,
           split between LGD, SGD, and NGD galaxies.}
  \tabcolsep 7pt
  \label{table_SN_morph}
   \begin{tabular}{lrrrrrr}
   \hline
  &\multicolumn{1}{c}{Sa}
  &\multicolumn{1}{c}{Sab}
  &\multicolumn{1}{c}{Sb}
  &\multicolumn{1}{c}{Sbc}
  &\multicolumn{1}{c}{Sc}
  &\multicolumn{1}{r}{All}\\
  \hline
  &\multicolumn{5}{c}{LGD (9, 12)}\\
  Ia & 2 & 4 & 7 & 21 & 27 & 61 \\
  Ibc & 1 & 0 & 2 & 14 & 19 & 36 \\
  II &0 & 1 & 13 & 22 & 61& 97 \\
  \\
  All & 3 & 5 & 22 & 57 & 107 & 194 \\
  \\
   &\multicolumn{5}{c}{SGD (5, 6)}\\
  Ia & 0 & 0 & 2 & 8 & 8 & 18 \\
  Ibc & 0 & 0 & 2 & 1 & 6 & 9 \\
  II & 0 & 0 & 3 & 7 & 22 & 32 \\
  \\
  All & 0 & 0 & 7 & 16 & 36 & 59 \\
  \\
   &\multicolumn{5}{c}{NGD (1-4)}\\
  Ia & 3 & 3 & 5 & 10 & 11 & 32 \\
  Ibc & 0 & 5 & 1 & 4 & 4 & 14 \\
  II & 1 & 1 & 4 & 9 & 19 & 34 \\
  \\
  All & 4 & 9 & 10 & 23 & 34 & 80 \\
  \hline
  \end{tabular}
  \parbox{\hsize}{\emph{Notes.} Among these 333 SNe,
                  there are only 23 uncertain (20 peculiar) classifications.
                  SNe of Type II include only 10 SNe IIb.
                  All Type IIn SNe are removed from the sample due to uncertainties in
                  their progenitor nature \citep[e.g.][]{2014MNRAS.441.2230H},
                  and often in their classification
                  \citep[e.g.][]{2013ApJS..207....3S,2018MNRAS.474..197P}.}
  \end{minipage}
\end{table}

In order to test our visual classification of spiral arms, the entire sample of SNe host galaxies was
independently classified by the first three authors of this paper.
By comparing these classifications, we determined that our ACs are 97 per cent reliable.
Following \citet[][]{2016MNRAS.459.3130A}, it is important to note that the most common
mis-classifications of ACs are from 2 or 3 to 4 (or vice versa), from 5 to 6 (or vice versa),
and from 9 to 12 (or vice versa).
Because we separated SNe host galaxies by their ACs into three broad classes:
LGD, SGD, and NGD, the possible mis-classification between them is negligible.

Of the sample galaxies, 56 are in common with galaxies for which ACs were determined by
\citet[][]{1987ApJ...314....3E} on the blue images of the Palomar Observatory Sky Survey
(POSS).\footnote{{\footnotesize For comparison of ACs,
another arm-classification by \citet[][]{2015ApJS..217...32B} might be used.
However, it is based on middle-infrared images (while we use the SDSS/optical images) and
another definition of broad ACs (flocculent: grouping 1-4 ACs, multi-arm: grouping 5-9 ACs, and GD: only AC 12),
which complicate the comparison.}}
A comparison of the ACs shows that about 65 per cent of the galaxies have the same broad classes.
On the other hand, about 25 per cent of objects change from NGD to SGD or from SGD to LGD (or vice versa).
The ACs change from NGD to LGD (or vice versa) only in about 10 per cent of the cases (6 individual galaxies).
In all the cases, the SDSS images have deeper exposure and better resolution that
the blue photographic plates of the POSS (in some cases, they are even overexposed
due to high surface brightness of the object).
Therefore, the SDSS based arm-classification seems to be more reliable and more structure is revealed.

The full database of 333 individual SNe
(SN designation, type, and offset from host galaxy nucleus) and
their 269 hosts (galaxy SDSS designation\footnote{{\footnotesize For the host galaxies
included in Table~\ref{dataRcorGal}, the PGC names are also available
in the database.}}, distance, morphological type,
$a/b$, PA, corrected $g$-band $D_{25}$, and AC)
is available in the online version (Supporting Information) of this article.

\section{Results}
\label{resdiscus}

To reveal the possible influence of DWs in discs of Sa--Sc galaxies
on the distribution and surface density of SNe,
we now study the deprojected and normalized galactocentric radii
of Type Ia and CC~SNe in discs of host galaxies with various ACs.

\subsection{The radial distribution and surface density}
\label{resdiscus_sub1}

In \citet[][]{2016MNRAS.456.2848H,2017MNRAS.471.1390H},
we already showed that in spiral galaxies all CC~SNe and
the overwhelming majority of Type Ia SNe belong to the disc,
rather than the bulge component.
Considering this observational fact, we adopt a simplified model where
all SNe are located on infinitely thin host discs and,
following \citet[][]{2009A&A...508.1259H},
we deproject the galactocentric radii of SNe ($R_{\rm SN}$) for the inclinations of these discs.
For each SN, we then normalize $R_{\rm SN}$ to the corresponding host galaxy optical radius,
i.e. $R_{25}=D_{25}/2$, to neutralize the greatly different linear (in kpc) sizes of
various hosts (as was shown in \citealt{2009A&A...508.1259H}).\footnote{{\footnotesize For the normalization,
one can suggest to use the SDSS scale lengths (exponential model fits) of galaxies.
However, our sample includes a large number of host galaxies with large angular sizes ($>100$~arcsec)
for which the SDSS fails in estimation of the model scale lengths due to the blending/defragmenting of
galaxies with large angular sizes (the scales are not reliable, this is well-known problem).
In \citet[][]{2012A&A...544A..81H}, we already commented about the SDSS model failure.
Thus, reliable scale lengths are not available for many galaxies of our sample.}}

\begin{table*}
  \centering
  \begin{minipage}{117mm}
  \caption{Comparison of the deprojected and normalized radial distributions of SNe ($\tilde{r}=R_{\rm SN}/R_{25}$)
           among different pairs of NGD, SGD, and LGD subsamples.
           The corresponding values for the inner-truncated disc
           ($\tilde{r} \ge 0.2$) are listed in parentheses.}
  \tabcolsep 5pt
  \label{RSNR25_KS_AD}
  \begin{tabular}{llrcllrcll}
  \hline
    \multicolumn{3}{c}{Subsample~1}&&\multicolumn{3}{c}{Subsample~2}&&\\
\multicolumn{1}{l}{Host}&\multicolumn{1}{c}{SN}&\multicolumn{1}{c}{$N_{\rm SN}$}&&\multicolumn{1}{l}{Host}&\multicolumn{1}{c}{SN}&\multicolumn{1}{c}{$N_{\rm SN}$}&\multicolumn{1}{c}{\,\, \,\,}&\multicolumn{1}{c}{$P_{\rm KS}$}&\multicolumn{1}{c}{$P_{\rm AD}$}\\
  \hline
  LGD&Ia&61 (50)&versus&NGD&Ia&32 (24)&&0.521 (0.497)&0.690 (0.708)\\
  LGD&Ia&61 (50)&versus&SGD&Ia&18 (16)&&0.761 (0.800)&0.505 (0.821)\\
  NGD&Ia&32 (24)&versus&SGD&Ia&18 (16)&&0.557 (0.641)&0.216 (0.671)\\
  \\
  LGD&CC&133 (111)&versus&NGD&CC&48 (40)&&0.087 (\textbf{0.048})&0.106 (\textbf{0.022})\\
  LGD&CC&133 (111)&versus&SGD&CC&41 (39)&&0.410 (0.096)&0.430 (0.125)\\
  NGD&CC&48 (40)&versus&SGD&CC&41 (39)&&0.080 (0.356)&0.108 (0.312)\\
  \\
  LGD&Ia&61 (50)&versus&LGD&CC&133 (111)&&0.720 (0.702)&0.719 (0.706)\\
  SGD&Ia&18 (16)&versus&SGD&CC&41 (39)&&0.834 (0.697)&0.862 (0.590)\\
  NGD&Ia&32 (24)&versus&NGD&CC&48 (40)&&0.545 (0.384)&0.284 (0.169)\\
  \hline
  \end{tabular}
  \parbox{\hsize}{\emph{Notes.} The probabilities from two-sample KS and AD tests ($P_{\rm KS}$ and $P_{\rm AD}$)
                  are calculated using the calibrations by \citet{Massey51} and \citet{Pettitt76}, respectively.
                  The statistically significant differences between the
                  distributions are highlighted in bold.}
  \end{minipage}
\end{table*}

\begin{figure}
\begin{center}$
\begin{array}{@{\hspace{0mm}}c@{\hspace{0mm}}}
\includegraphics[width=1\hsize]{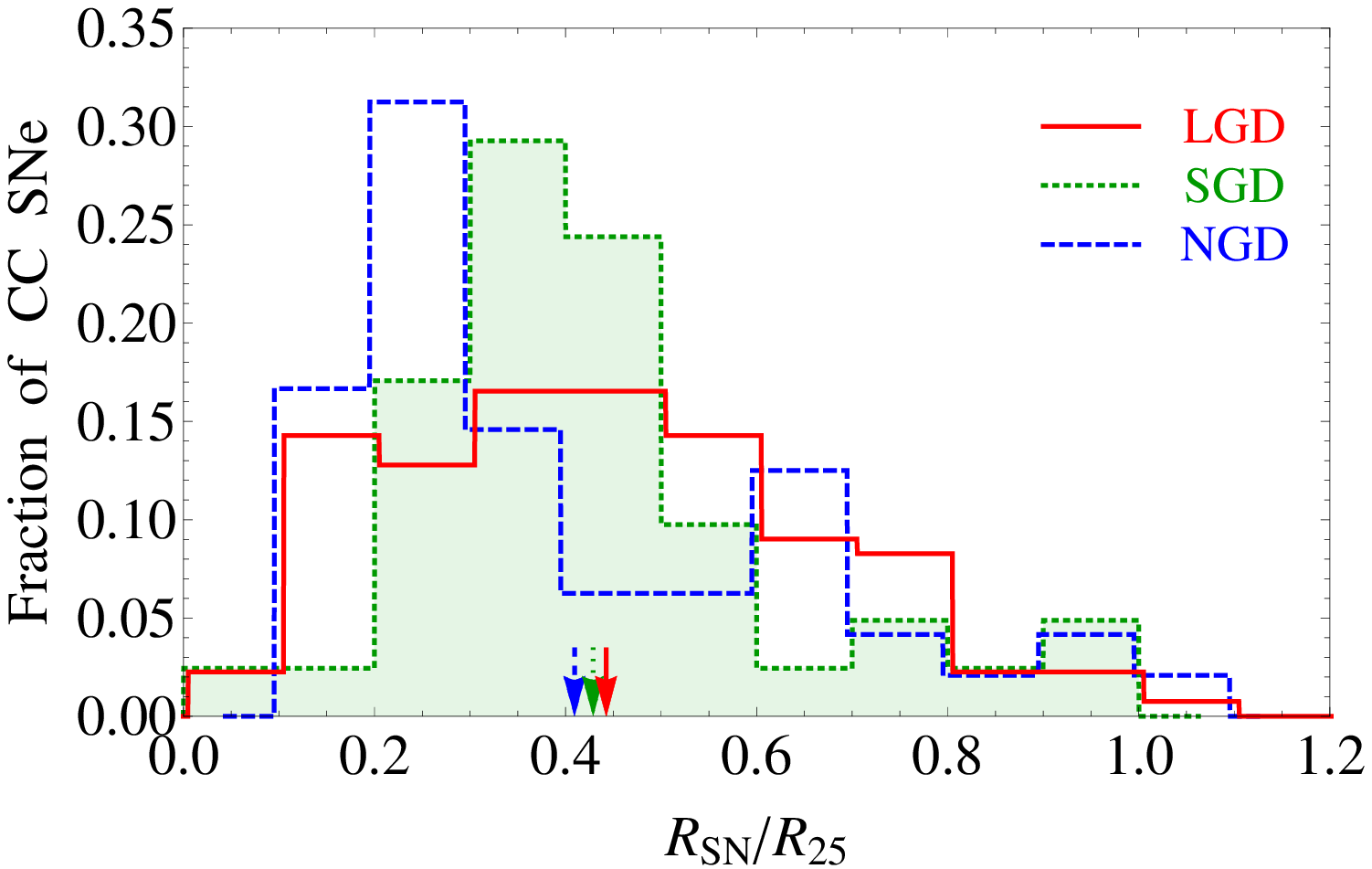}
\end{array}$
\end{center}
\begin{center}$
\begin{array}{@{\hspace{0mm}}c@{\hspace{0mm}}}
\includegraphics[width=1\hsize]{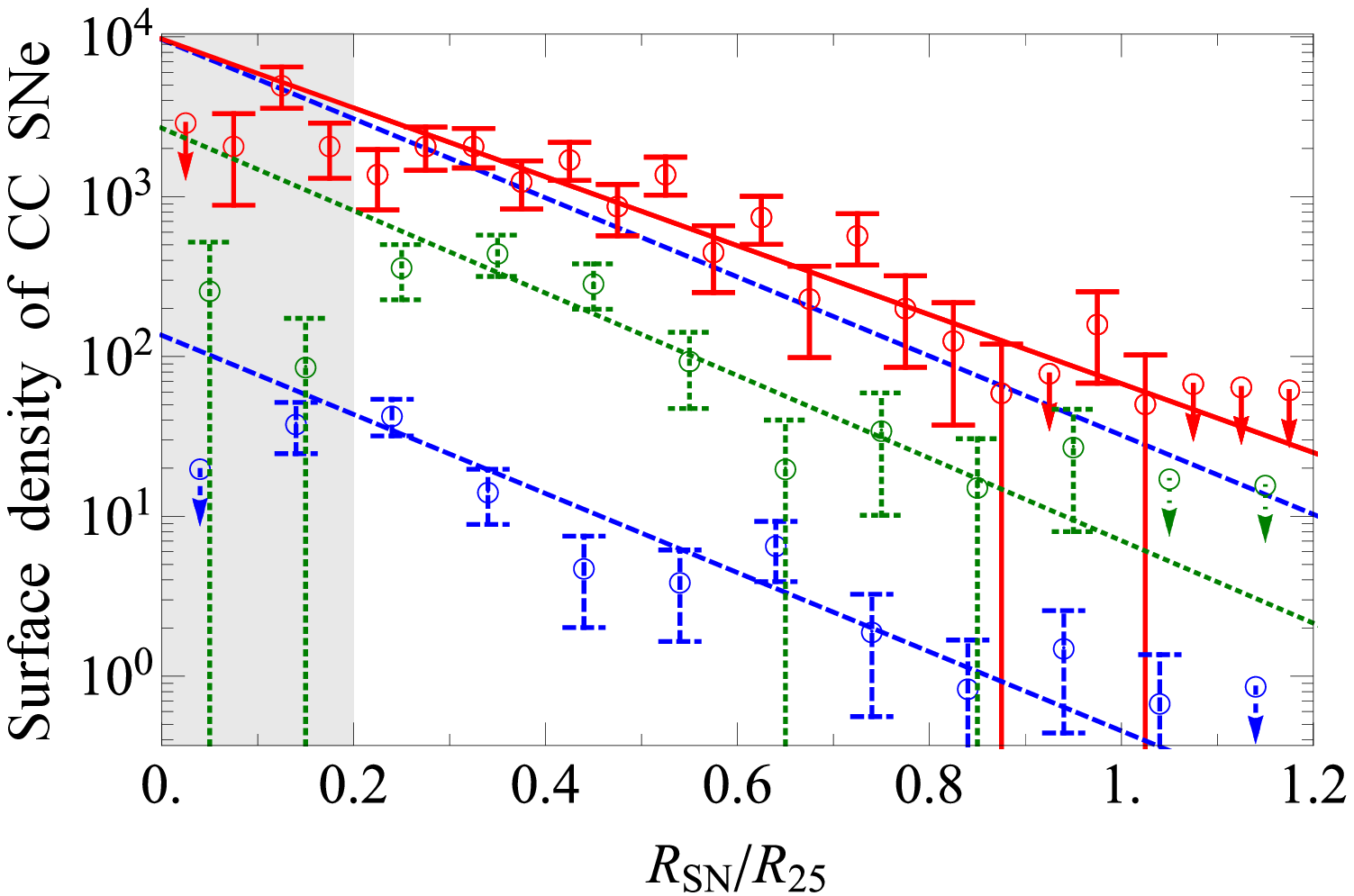}
\end{array}$
\end{center}
\begin{center}$
\begin{array}{@{\hspace{0mm}}c@{\hspace{0mm}}}
\includegraphics[width=1\hsize]{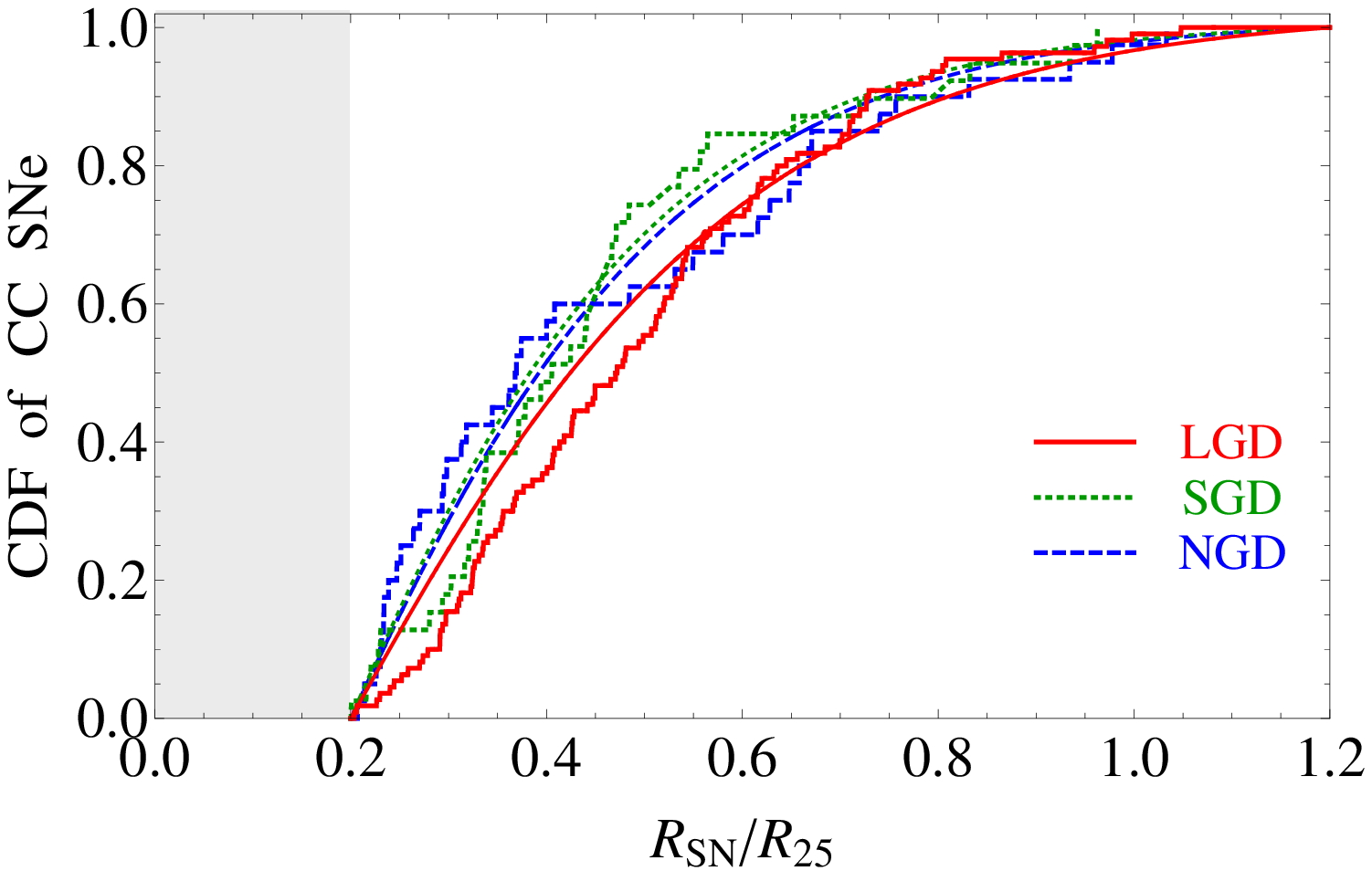}
\end{array}$
\end{center}
\caption{\emph{Upper panel}: distributions of deprojected and normalized galactocentric radii ($\tilde{r}=R_{\rm SN}/R_{25}$) of
         CC~SNe in LGD (red solid), SGD (green dotted), and NGD (blue dashed) host galaxies.
         The mean values of the distributions are shown by arrows.
         \emph{Middle panel}: surface density distributions (with arbitrary normalization)
         of CC~SNe in the mentioned hosts.
         For better visualization, thanks to more data points, the bin size of distribution in LGD
         galaxies is 0.05, in units of $R_{25}$, while for the other subsamples the bin size is 0.1.
         The error bars assume a Poisson distribution. The upper-limits of surface density
         (with $+1$ SN if none is found)
         are represented by down arrows. The fitted exponential surface density profiles are estimated for
         the inner-truncated discs (outside the shaded area).
         For better visibility, the distributions and profiles are shifted vertically sorted by increasing
         the mean $\tilde{r}$ as one moves upwards, and also slightly shifted horizontally.
         To visually compare the distribution of CC~SNe in LGD hosts with the fitted profile in NGD galaxies,
         the latter is also positioned with the central surface density
         matched with that in LGD hosts.
         \emph{Bottom panel}: inner-truncated cumulative distributions of SN radii
         and their best-fitting exponential CDFs.}
\label{HistSDFSurfd}
\end{figure}

In Table~\ref{RSNR25_KS_AD}, using the two-sample Kolmogorov--Smirnov (KS) and
Anderson--Darling (AD) tests,\footnote{{\footnotesize The null hypothesis
for the two-sample nonparametric KS (or AD) test is that the two distributions
being compared are drawn from the same parent population, and the alternative hypothesis
that they are not. Traditionally, we chose the threshold of 5 per cent for
significance levels ($P$-values) of the tests.
The AD test detects differences better than the KS test and generally requires less data
to reach sufficient statistical power \cite[][]{Engmann+11}.}}
we compare the deprojected and normalized ($\tilde{r}=R_{\rm SN}/R_{25}$)
radial distributions of Type Ia and CC~SNe in different pairs of NGD, SGD, and LGD subsamples.
From the $P$-values in Table~\ref{RSNR25_KS_AD}, we see no statistically significant differences
between the radial distributions of SNe in various subsamples.
However, when we compare the inner truncated radial distributions
($\tilde{r} \geq 0.2$; shown in brackets),
a significant difference appears for CC~SNe in LGD versus NGD hosts.
The upper panel of Fig.~\ref{HistSDFSurfd} presents the histograms of radii
of CC~SNe.
From these histograms,
we see that the radial distribution of CC~SNe in NGD subsample is
concentrated to the centre of galaxies with a relatively narrow peak and
fast decline in the outer disc.
In contrast, the distribution of CC~SNe in LGD galaxies has a broader peak,
shifted to the outer region of the discs, with a somewhat slower decline.
The radial distribution of SNe in SGD hosts appears to be intermediate
between those in NGD and LGD galaxies.

The inner truncation of the radial distribution of SNe, especially for CC ones,
is crucial because of several important effects.
The observed numbers of SNe at $\tilde{r}\lesssim0.2$ indicate that because of
high surface brightness of galactic nuclei and imperfect reduction of astronomical images
it is difficult to discover objects at or near the centre of galaxies,
even for nearby ones \citep[e.g.][]{2011MNRAS.412.1419L}.
On the other hand, dust extinction in host galaxy disc,
particularly in the nuclear region \cite[e.g.][]{2015MNRAS.446.3768H},
can affect the radial distributions of SNe \cite[e.g.][]{1997ApJ...483L..29W,1998ApJ...502..177H}.
Since CC~SNe have peak luminosities that are $\sim2$ magnitudes lower than
do SNe~Ia \citep[e.g.][]{2002AJ....123..745R}, CC~SNe
are more strongly affected by these effects than are Type Ia SNe.

\begin{table*}
  \centering
  \begin{minipage}{123mm}
  \caption{Consistency of global ($\tilde{r}\ge0$) and
           inner-truncated ($\tilde{r}\ge0.2$) SN distributions with exponential surface
           density models in different subsamples of host galaxies.}
  \tabcolsep 5pt
  \label{tableGDNGDexp}
  \begin{tabular}{llcrccccrccc}
  \hline
  &&&
  \multicolumn{4}{c}{$\tilde{r}\ge0$}&
  \multicolumn{1}{c}{}&
  \multicolumn{4}{c}{$\tilde{r}\ge0.2$}\\
  \multicolumn{1}{l}{Host}&
  \multicolumn{1}{c}{SN}&\multicolumn{1}{c}{\,\, \,\,}&
  \multicolumn{1}{c}{$N_{\rm SN}$}&
  \multicolumn{1}{c}{$P_{\rm KS}$}&
  \multicolumn{1}{c}{$P_{\rm AD}$}&
  \multicolumn{1}{c}{$\tilde{h}_{\rm SN}$}&\multicolumn{1}{c}{\,\, \,\,}&
  \multicolumn{1}{c}{$N_{\rm SN}$}&
  \multicolumn{1}{c}{$P_{\rm KS}$}&
  \multicolumn{1}{c}{$P_{\rm AD}$}&
  \multicolumn{1}{c}{$\tilde{h}_{\rm SN}$}\\
  \multicolumn{1}{l}{(1)}&\multicolumn{1}{c}{(2)}&&
  \multicolumn{1}{c}{(3)}&\multicolumn{1}{c}{(4)}&
  \multicolumn{1}{c}{(5)}&\multicolumn{1}{c}{(6)}&&
  \multicolumn{1}{c}{(7)}&\multicolumn{1}{c}{(8)}&
  \multicolumn{1}{c}{(9)}&\multicolumn{1}{c}{(10)}\\
  \hline
  All&Ia&&111&0.141&0.148&$0.21\pm0.01$&&90&0.272&0.333&$0.20\pm0.01$\\
  LGD&Ia&&61&0.730&0.514&$0.22\pm0.02$&&50&0.886&0.791&$0.20\pm0.02$\\
  SGD&Ia&&18&0.318&0.290&$0.24\pm0.03$&&16&0.737&0.506&$0.21\pm0.03$\\
  NGD&Ia&&32&0.557&0.489&$0.19\pm0.02$&&24&0.449&0.379&$0.18\pm0.02$\\
  \\
  All&CC&&222&\textbf{0.005}&\textbf{0.002}&$0.22\pm0.01$&&190&0.117&0.172&$0.19\pm0.01$\\
  LGD&CC&&133&\textbf{0.017}&\textbf{0.018}&$0.22\pm0.01$&&111&0.070&\textbf{0.043}&$0.20\pm0.01$\\
  SGD&CC&&41&\textbf{0.023}&\textbf{0.035}&$0.21\pm0.01$&&39&0.349&0.440&$0.17\pm0.02$\\
  NGD&CC&&48&0.191&0.180&$0.20\pm0.02$&&40&0.579&0.407&$0.18\pm0.02$\\
  \hline
  \end{tabular}
  \parbox{\hsize}{
    \emph{Notes.} Columns~1 and 2 give the subsample;
    Col.~3 is the number of SNe in the subsample;
    Cols.~4 and 5 are the $P_{\rm KS}$ and $P_{\rm AD}$ probabilities from one-sample KS and AD tests,
    respectively, that the global ($\tilde{r}\ge0$) distribution of SNe is drawn from
    the best-fitting exponential surface density profile;
    Col.~6 is the maximum likelihood value of $\tilde h_{\rm SN} = h_{\rm SN}/R_{25}$
    with bootstrapped error (repeated $10^3$ times);
    Cols.~7--10 are respectively the same as Cols.~3--6,
    but for the inner-truncated ($\tilde{r}\ge0.2$) distribution.
    The $P_{\rm KS}$ and $P_{\rm AD}$ are calculated using the calibrations by
    \citet{Massey51} and \citet{1986gft..book.....D}, respectively.
    The statistically significant deviations from an exponential law are highlighted in bold.}
  \end{minipage}
\end{table*}

In \citet[][]{2016MNRAS.456.2848H}, we already demonstrated that in the central regions of
unbarred spiral galaxies the surface densities of SNe show a drop, significantly for CC~SNe
(see also in the middle panel of Fig.~\ref{HistSDFSurfd}),
in comparison with the exponential surface density profiles of the parent populations
\cite[see also][]{1997AJ....113..197V,2010MNRAS.405.2529W}.
We list, in columns~4 and 5 of Table~\ref{tableGDNGDexp},
for different subsamples of the present study,
the $P_{\rm KS}$ and $P_{\rm AD}$ probabilities from one-sample KS and AD tests, respectively,
that the distributions of SNe are drawn from the best-fitting exponential surface density profiles.
We obtain $\Sigma^{\rm SN}(\tilde{r})=\Sigma_0^{\rm SN} \exp(-\tilde{r}/\tilde{h}_{\rm SN})$
profiles using the maximum likelihood estimation (MLE) method,
where $\tilde{h}_{\rm SN}$ is the scale length of the distribution (column~6 of Table~\ref{tableGDNGDexp})
and $\Sigma_0^{\rm SN}$ is the central surface density of SNe.
The $P$-values in Table~\ref{tableGDNGDexp} show that the global ($\tilde{r}\ge0$) distributions of Type Ia SNe
in different subsamples are consistent with the exponential profiles.
However, the surface density distributions of CC~SNe are not consistent with
the exponential profiles in all subsamples of host galaxies, except the NGD hosts.

To exclude the selection effects at the centres of host galaxies,
we repeat our procedure for $\tilde{r}\ge0.2$ range
(see columns~7--10 in Table~\ref{tableGDNGDexp}).
Now, with only one exception, the surface density distributions of Type Ia and CC~SNe
in different subsamples are consistent with the exponential profiles.
The inner-truncated scale lengths are in agreement with those in
\citet[][]{2016MNRAS.456.2848H}: using nearby low-inclined early-type spiral galaxies
(unbarred Sa--Sbc, without splitting the sample according to ACs)
we found $\tilde{h}_{\rm SN}^{\rm Ia}=0.21\pm0.03$ and
$\tilde{h}_{\rm SN}^{\rm CC}=0.17\pm0.03$ in the SDSS $g$-band.

Only the surface density distribution of CC~SNe in LGD galaxies is inconsistent with
an inner-truncated exponential profile (as seen in Table~\ref{tableGDNGDexp} for
the AD statistic but only very marginally so in the KS statistic).
From the middle panel of Fig.~\ref{HistSDFSurfd},
we see that the surface density
is marginally higher than the best-fitting exponential profile at
$0.4\lesssim\tilde{r}\lesssim0.7$.
The inconsistency becomes more evident if we compare the distribution of CC~SNe
in LGD galaxies with the inner-truncated exponential profile
with the scale length of CC~SNe in NGD galaxies
($P_{\rm KS}=0.005$ and $P_{\rm AD}=0.001$).
For the visualization, the latter (upper blue dashed line
in the middle panel of Fig.~\ref{HistSDFSurfd}) is also scaled according to
the central surface density of the profile in LGD hosts.
The bottom panel of Fig.~\ref{HistSDFSurfd} shows the cumulative distributions
of CC~SN radii with their best-fitting exponential cumulative distribution functions (CDFs).

\section{Interpretation within the framework of density wave theory}
\label{DWTInt}

In this section, we interpret the results above
in the context of triggered massive star formation by the DWs in GD galaxies,
especially in LGD hosts
\citep[e.g.][]{1990ApJ...349..497C,2002MNRAS.337.1113S,2013AA...560A..59C}.

\begin{figure}
\begin{center}
\includegraphics[width=1\hsize]{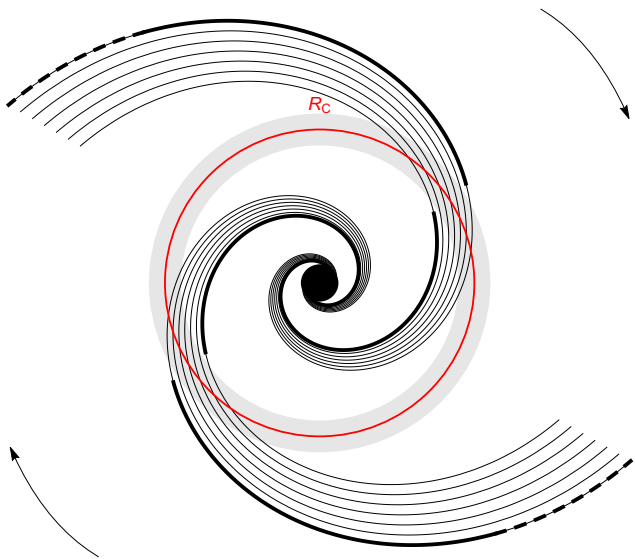}
\end{center}
\caption{The scheme of massive star formation triggering by DWs
           in a model of a GD galaxy with two logarithmic spiral arms.
           The direction of galaxy rotation is illustrated by arrows.
           The shock fronts of spiral arms are displayed with thick black curves.
           The corotation region and radius $R_{\rm C}$ are represented by
           a thick gray ring and red solid circle, respectively.
           At large radii ($\gtrsim R_{25}$),
           the impact of DWs is expected to be weak
           (shock fronts are presented by thick dashed curves).}
\label{armmodlab}
\end{figure}

In a simple model of GD host galaxies (Fig.~\ref{armmodlab}),
we assume that the spiral pattern rotates with a constant angular velocity,
while the gas and stars have differential rotation,
and a corotation radius/region ($R_{\rm C}$)
exists where these two angular velocities are equal.
Inside the corotation radius, the disc rotates faster than the spiral arm pattern,
and therefore massive star formation triggering is expected in a shock front around the inner edges of arms
(thick black curves inside red solid circle in Fig.~\ref{armmodlab},
see also fig.~9 of \citealt[][]{2016MNRAS.459.3130A}),
as originally proposed by \citet[][]{1969ApJ...158..123R}.
On the contrary, outside the corotation radius, the arm pattern rotates faster than the disc.
Therefore, gas and stars are caught up by the spiral arms.
In this case, star formation is expected to be triggered in a shock front
around the outer edges of arms (thick black curves outside red solid circle in Fig.~\ref{armmodlab}).
Indeed, in \citet[][]{2016MNRAS.459.3130A}, we already showed that the distribution of CC~SNe
(explosions of young short-lived massive stars)
relative to the SDSS $g$-band peaks of spiral arms depends on the galactocentric radial range.
In particular, the locations of CC~SNe are shifted to the inner and outer edges
of the spiral arms inside and outside the mean corotation radius
($\langle R_{\rm C}/R_{25}\rangle \approx0.45$) of LGD galaxies, respectively.
For Type Ia SNe (explosions of less-massive and longer-lived stars),
the distribution relative to spiral arms
showed no significant dependence on galactocentric radii.

In the corotation region (thick gray ring in Fig.~\ref{armmodlab}) where the
stars and gas rotate at the same velocity as the spiral pattern,
the triggering of star formation is not expected, given the absence of spiral
shocks.
Mainly, the gravitation instability is responsible for the star formation in
this region (as in the entire disc of a NGD galaxy).
Due to absence of star formation triggering spiral shocks in
the corotation region \citep[e.g.][]{2013AA...560A..59C},
the surface density of CC~SNe should show a drop around $R_{\rm C}$ in GD galaxies.
At the same time, at large radii ($\gtrsim R_{25}$) the DWs are expected to fade
\citep[e.g.][]{1992ApJS...79...37E}.
Therefore, at large radii, star formation triggered by shock fronts at the outer edges of arms
should be not significant (thick black dashed curves in Fig.~\ref{armmodlab}).

\begin{table*}
  \centering
  \begin{minipage}{167mm}
  \caption{Available corotation radii of our LGD and SGD host
    galaxies.}
  \tabcolsep 5pt
  \label{dataRcorGal}
  \begin{tabular}{lrrrccl}
  \hline
 \multicolumn{1}{l}{Host name}
&\multicolumn{1}{c}{AC}
&\multicolumn{1}{c}{$N_{\rm Ia}$}
&\multicolumn{1}{c}{$N_{\rm CC}$}
&\multicolumn{1}{c}{$R_{\rm C}/R_{25}$}
&\multicolumn{1}{c}{$R_{\rm C}/R_{25}$}
&\multicolumn{1}{c}{References}\\
 \multicolumn{1}{c}{(1)}&\multicolumn{1}{c}{(2)}&
\multicolumn{1}{c}{(3)}&\multicolumn{1}{c}{(4)}&
\multicolumn{1}{c}{(5)}&\multicolumn{1}{c}{(6)}&
\multicolumn{1}{c}{(7)}\\
  \hline
    PGC043118&12&1&0&$0.33\pm0.05$&&\citet{2014AA...562A.121C}\\
    PGC040153&12&1&1&$0.30\pm0.05$&&\citet{1997ApJ...479..723C,2014AA...562A.121C}\\
    PGC038068&12&0&3&$0.50\pm0.08$&&\citet{2008MNRAS.388.1803R,2009ApJS..182..559B}\\
    PGC030087&12&0&4&$0.54\pm0.13$&&\citet{2008AJ....136.2872T}\\
    PGC024531&12&0&1&$0.87\pm0.11$&&\citet{2007AA...474...43V,2014ApJS..210....2F}\\
    PGC007525&12&0&2&$0.30\pm0.06$&&\citet{2007AA...474...43V}\\
    PGC005974&12&0&3&$0.34\pm0.09$&&\citet{1992ApJS...79...37E,2009ApJ...697.1870E,2013AA...560A..59C}\\
    PGC054018&9&0&1&$0.40\pm0.04$&&\citet{2014ApJS..210....2F}\\
    PGC050063&9&1&3&$0.21\pm0.03$&$0.45\pm0.12$&\citet{1992ApJS...79...37E,1997ApJ...481..169W,2013AA...560A..59C}\\
    PGC042833&9&0&2&$0.37\pm0.04$&$0.57\pm0.05$&\citet{2009ApJS..182..559B,2014ApJS..210....2F}\\
    PGC039578&9&0&4&$0.34\pm0.06$&$0.57\pm0.07$&\citet{1992ApJS...79...37E,1996ApJ...460..651G,2009ApJS..182..559B}\\
    PGC038618&9&0&1&$0.30\pm0.01$&$0.54\pm0.06$&\citet{2009ApJS..182..559B}\\
    PGC037845&9&0&1&$0.21\pm0.06$&$0.40\pm0.06$&\citet{2009ApJS..182..559B}\\
    PGC037229&9&0&4&$0.46\pm0.08$&&\citet{1992ApJS...79...37E,2009ApJS..182..559B}\\
    PGC036789&9&0&1&$0.22\pm0.06$&&\citet{2014AA...562A.121C}\\
    PGC036243&9&0&2&$0.45\pm0.13$&&\citet{2003ApJ...586..143K,2009ApJS..182..559B}\\
    PGC034767&9&0&3&$0.28\pm0.03$&&\citet{2001MNRAS.323..651F}\\
    PGC032614&9&0&2&$0.69\pm0.02$&$0.83\pm0.04$&\citet{2014ApJS..210....2F}\\
    PGC031968&9&0&1&$0.26\pm0.02$&&\citet{2014ApJS..210....2F}\\
    PGC027074&9&0&1&$0.30\pm0.06$&&\citet{2014AA...562A.121C}\\
    PGC024111&9&1&1&$0.65\pm0.06$&&\citet{2014AA...562A.121C}\\
    PGC022279&9&0&1&$0.16\pm0.06$&&\citet{2007AA...474...43V}\\
    PGC002246&9&0&1&$0.14\pm0.06$&$0.57\pm0.06$&\citet{2007AA...474...43V}\\
    PGC002081&9&0&1&$0.38\pm0.05$&&\citet{1992ApJS...79...37E,1997AA...317..405S}\\
    PGC038031&6&1&0&$0.22\pm0.03$&$0.42\pm0.02$&\citet{2014ApJS..210....2F,2014AA...562A.121C}\\
    PGC027723&6&1&0&$0.17\pm0.06$&$0.44\pm0.06$&\citet{2014AA...562A.121C}\\
    PGC012626&6&2&0&$0.48\pm0.03$&&\citet{2009ApJS..182..559B}\\
    PGC035594&5&0&1&$0.32\pm0.06$&&\citet{2014ApJS..210....2F}\\
    PGC034836&5&0&2&$0.12\pm0.06$&$0.58\pm0.06$&\citet{2009ApJS..182..559B}\\
    PGC030010&5&0&1&$0.17\pm0.06$&$0.41\pm0.06$&\citet{2014AA...562A.121C}\\
  \hline
  \end{tabular}
  \parbox{\hsize}{\emph{Notes.} Column~1 is the host galaxy PGC name;
                  Col.~2 is the galaxy AC (see Subsection~\ref{sample2});
                  Cols.~3 and 4 are the numbers of Type Ia and CC~SNe in the galaxy;
                  Cols.~5 and 6 are the normalized corotation radii of the galaxy;
                  Col.~7 is the references of corotation radii.
                  The $R_{\rm C}/R_{25}$ values are calculated using the $R_{\rm C}$ in arcsec
                  from the mentioned references and the galaxy $R_{25}$ in the SDSS $g$-band
                  (see Subsection~\ref{sample1}).
                  When more than one references are available for the same corotation region and
                  the reported radii are matched within the errors, we list their mean values.
                  Nuclear and circumnuclear corotation radii (coincided with star-forming rings/ovals),
                  as well as those with uncertain (very weak/noisy) estimation are not selected from the references.}
  \end{minipage}
\end{table*}

To study the distribution of SNe relative to $R_{\rm C}$ of hosts,
we carried out an extensive literature search for corotation radii of our SGD and LGD galaxies.
Only 30 nearby host galaxies (${\lesssim {\rm 80~Mpc}}$) with 8 Type Ia and 48 CC~SNe
have available corotation radii (Table~\ref{dataRcorGal}).
These radii were estimated using different methods.
For example, \citet{1992ApJS...79...37E} found clear evidences for the corotation radii
in gas-rich galaxies, in the form of sharp endpoints to star formation ridges and dust lanes in GD spirals.
\citet{2007AA...474...43V} used Fourier analysis and focused on the modes of the spiral arms,
computing the torques between the gas and newly formed stars (H$\alpha$ emission),
and the bulk of the optical matter ($r$-band), which can be used to locate the corotation regions.
\citet{2009ApJS..182..559B} used the potential-density phase-shift method on
deprojected $H$-band images to locate the corotation radii for a large number of spiral galaxies.
\citet{2014ApJS..210....2F} used the changes in direction of the radial component of the in-plane velocities,
using the emission in H$\alpha$, at the resonance radii to find corotations in disc galaxies.
For more details of these and other methods, the reader is referred to the original papers
mentioned in Table~\ref{dataRcorGal}.
Farther in our study, we use these corotation radii normalized to the optical radii of host galaxies
in the SDSS $g$-band, i.e. $R_{\rm C}/R_{25}$.

In Table~\ref{dataRcorGal}, it can be seen that for some individual galaxies more than one corotation radius is found.
This is not unexpected because real spiral galaxies are more complex physical objects
in comparison with the simple model presented in Fig.~\ref{armmodlab}.
In some galaxies, single pattern velocities and single corotation radii are observed,
while in other systems multiple spiral patterns with different velocities and resonant coupling \citep[e.g.][]{2009ApJ...702..277M},
and therefore multiple corotation radii are discovered \citep[e.g.][]{2009ApJS..182..559B,2014ApJS..210....2F}.
In particular, \citet{2009ApJS..182..559B} found that GD galaxies have on average 2-3 corotation radii,
except for exceptionally strong GD spirals (AC$=$12), which mostly have a
single corotation radius.
This is in agreement with our ACs of SNe host galaxies and collected corotation radii in Table~\ref{dataRcorGal}.

\begin{figure*}
  \begin{center}
  \includegraphics[width=\hsize]{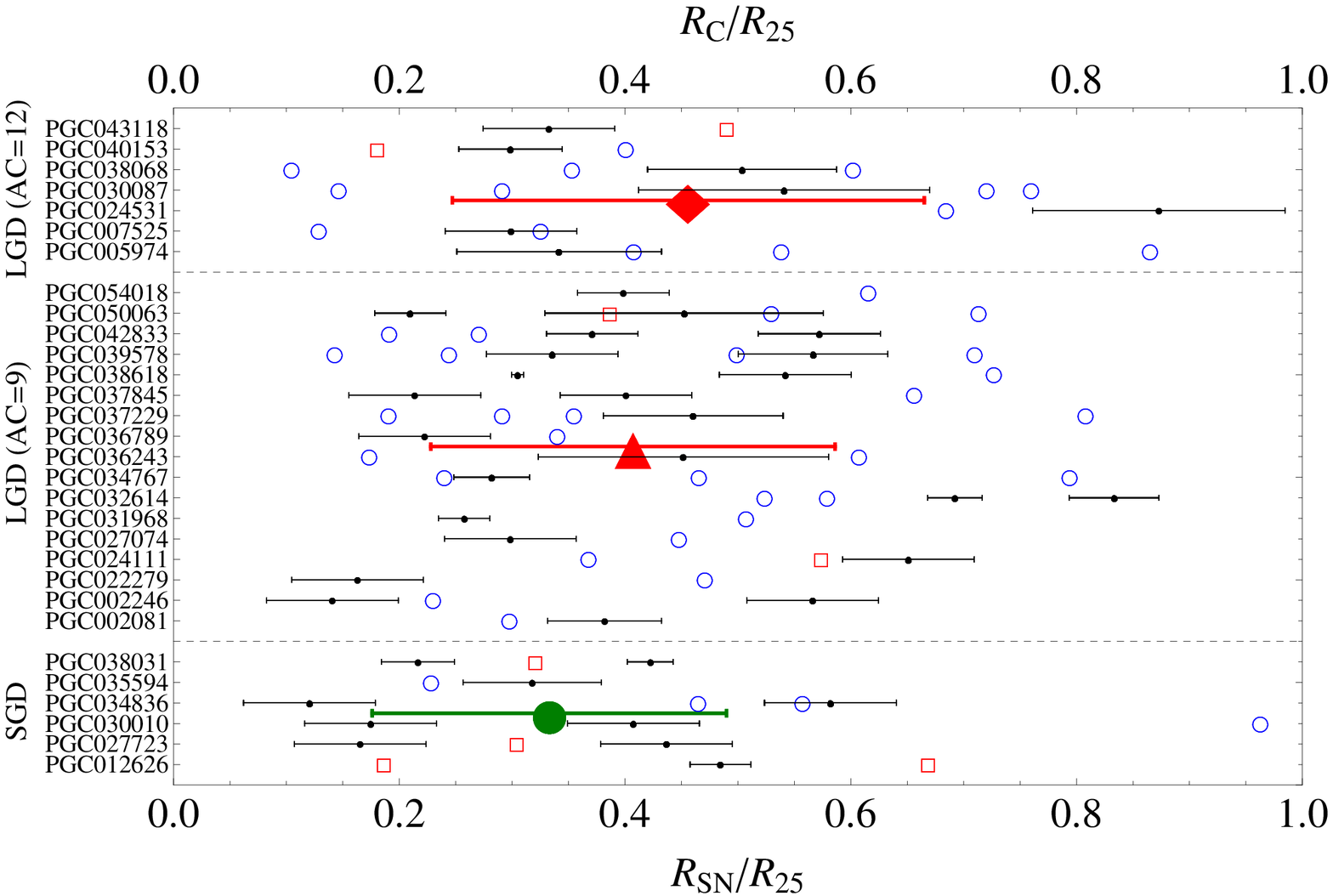}
  \end{center}
  \caption{Galactocentric positions of normalized corotation radii (black points) and their errors
           for 30 host galaxies of Table~\ref{dataRcorGal}.
           SGD (AC$=$5 and 6), LGD (AC$=$9), and LGD (AC$=$12) galaxies are separated by horizontal dashed lines.
           The filled diamond, triangle, and circle are the corresponding mean values of
           the corotation radii (with their standard deviations).
           For each host galaxy, galactocentric positions of Type Ia (red empty squares) and
           CC (blue empty circles) SNe are also presented.
           In PGC~050063, one of the CC~SNe is located at $R_{\rm SN}/R_{25}=1.59$
           and not shown in the plot.}
  \label{RcR25}
\end{figure*}

In Fig.~\ref{RcR25}, we present the galactocentric $R_{\rm C}/R_{25}$ positions
for 30 host galaxies of Table~\ref{dataRcorGal}, separated according to their ACs:
6 SGD (AC$=$5 and 6), 17 LGD (AC$=$9), and 7 LGD (AC$=$12) galaxies.
Here, we separate LGD host galaxies between two ACs in order to check
possible differences between the distributions and the mean values of their corotation radii.
A similar separation is impossible for SGD galaxies due to the small size of
this subsample (see Table~\ref{dataRcorGal}).
In Fig.~\ref{RcR25}, we also show the galactocentric $R_{\rm SN}/R_{25}$
positions of SNe for each host galaxy.

\begin{figure}
  \centering
  \includegraphics[width=1\hsize]{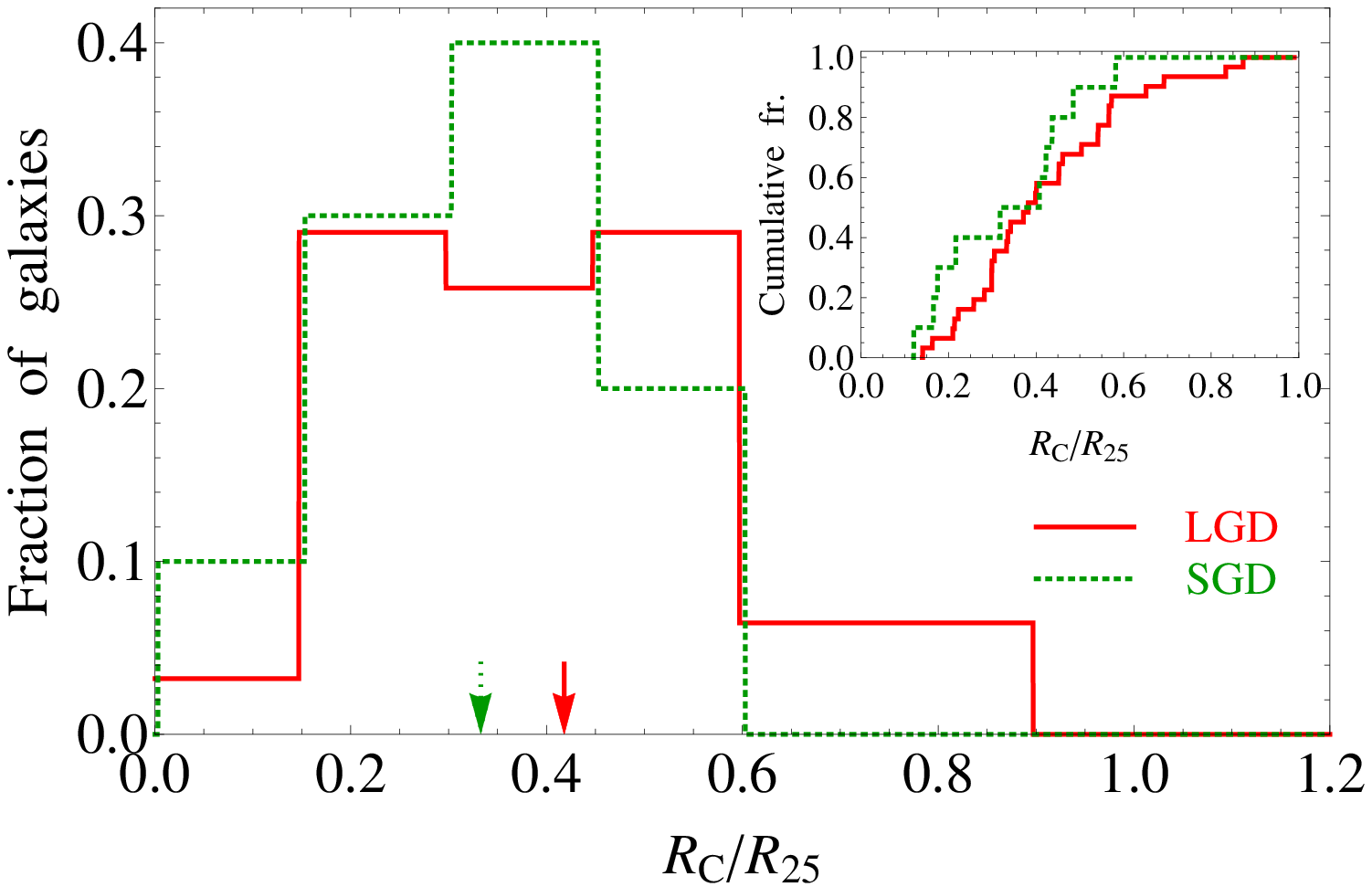}
  \caption{Histograms and cumulative distributions (inset) of $R_{\rm C}/R_{25}$ values of
           LGD (red solid) and SGD (green dotted) galaxies.
           The mean values are shown by arrows.}
  \label{RCR25histcum}
\end{figure}

The mean values of normalized corotation radii and the standard deviations are
0.33$\pm$0.16, $0.41\pm$0.18, and $0.46\pm0.21$ for
SGD, LGD with AC$=$9 and AC$=$12 galaxies, respectively.
For the united LGD (AC$=$9 and 12) subsample, the normalized corotation radius is $0.42\pm0.18$.
Meanwhile, the two-sample KS and AD tests show that the difference between the distributions
of $R_{\rm C}/R_{25}$ values in LGD (AC$=$9 and 12) and SGD galaxies is statistically not significant
($P_{\rm KS}=0.550$ and $P_{\rm AD}=0.312$).
The same is valid when comparing the $R_{\rm C}/R_{25}$ distributions in LGD (AC$=$12) and SGD galaxies
($P_{\rm KS}=0.433$ and $P_{\rm AD}=0.268$).
Therefore, further in our study we do not separate the LGD subsample.
Fig.~\ref{RCR25histcum} shows the histograms and cumulative distributions of
$R_{\rm C}/R_{25}$ values of LGD and SGD galaxies.
Also, it is important to note, that the
$\langle R_{\rm C}/R_{25}\rangle$ value for LGD galaxies is in good agreement with
that ($\approx0.45$) adopted in our previous study \citep{2016MNRAS.459.3130A}.

\begin{figure}
  \centering
  \includegraphics[width=1\hsize]{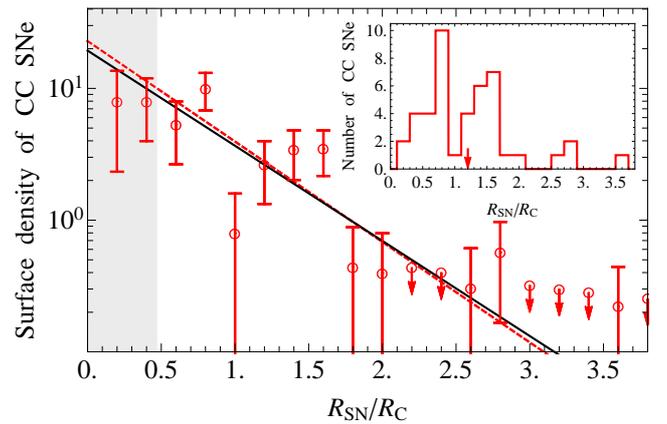}
  \caption{Surface density profile of CC~SNe (with arbitrary normalization)
    in LGD host galaxies.
    The error bars assume a Poisson distribution.
    The upper-limits of surface density (with $+1$ SN if none is found)
    are represented by down arrows.
    The black solid and red dashed lines are the best
    maximum-likelihood fits of global and inner-truncated
    (from 0.48 corotation radii outwards to avoid the obscured inner region [grey shaded])
    exponential surface density models, respectively.
    The inset presents the histogram of SN radii (the mean value is shown by arrow).}
  \label{surfdens}
\end{figure}
\begin{figure}
  \centering
  \includegraphics[width=1\hsize]{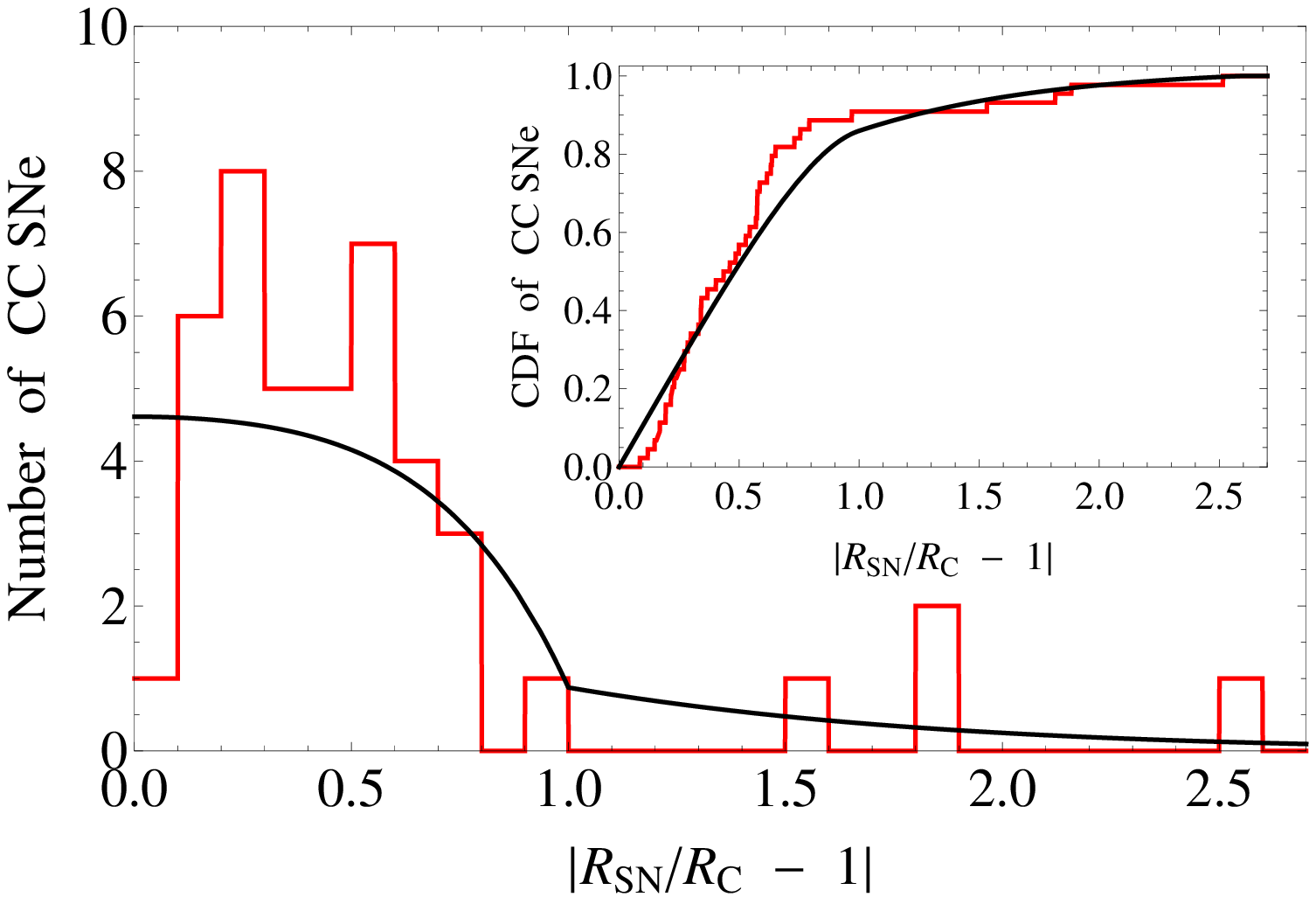}
  \caption{Differential distribution of the distances of CC~SNe to nearest corotation radius in
    the global disc of LGD galaxies (normalized to the corotation radius).
    The inset presents the CDF of the distances. The black curves indicate the
    distribution of normalized distances to corotation expected for the
    best-fitting exponential surface density model (scale length of
    $0.60\,R_{\rm C}$), using eqs.~(\ref{PDFD}), (\ref{PDFD2}), (\ref{CDFD}), and (\ref{CDFD2}).}
  \label{distCorot}
\end{figure}

To check the possible impact of DWs on the distribution of SNe
(as schematically presented in Fig.~\ref{armmodlab}),
we now normalize the SN radii to the corresponding corotation radii of host galaxies.
When a host galaxy has two corotation radii in Table~\ref{dataRcorGal},
we use a proximity criterion, selecting only the value of $R_{\rm C}$ that is
closest to the value of $R_{\rm SN}$.
For LGD host galaxies, Fig.~\ref{surfdens} displays the histogram and surface density
of 44 CC~SNe positions in units of the corotation radii ($R_{\rm SN}/R_{\rm C}$).
The surface density of CC~SNe is consistent with the best-fitting
global ($P_{\rm KS}=0.600$, $P_{\rm AD}=0.463$)
and inner-truncated ($P_{\rm KS}=0.457$, $P_{\rm AD}=0.526$) exponential profiles
with the MLE scale lengths of $(0.60\pm0.04)\,R_{\rm C}$ and $(0.57\pm0.05)\,R_{\rm C}$, respectively.
However, the figure indicates a strong dip at the corotation radius, and excess surface
densities of CC~SNe at $\simeq 0.8$ and $1.5\,R_{\rm C}$.

Since the lifetime of massive progenitors of CC~SNe is significantly short,
their explosion sites, on average, coincide with the birthplace.
Therefore, the prominently high surface density of CC~SNe
in comparison with the best-fitting exponential profile around
the mentioned radii, inside and outside the corotation region,
can be considered as a plausible indicator of
triggered massive star formation by the DWs in LGD host galaxies.
These results are in agreement with those of \citet{2013AA...560A..59C},
who found clear evidence of massive star formation triggering in
the sense of a high density of H~{\footnotesize II} regions at the fixed radii,
avoiding the corotation region, created after the passage of the arm material
through the DW in some GD galaxies
\citep[see also][]{1990ApJ...349..497C,2002MNRAS.337.1113S}.

Considering that the different LGD host galaxies have various corotation radii
(see Table~\ref{dataRcorGal} and Fig.~\ref{RcR25})
distributed around the mean value of
$\langle R_{\rm C}/R_{25}\rangle=0.42\pm0.18$ (see Fig.~\ref{RCR25histcum}),
the radii of triggered star formation by DWs should be blurred
within a radial region including $\sim0.4$ to $\sim0.7$ range in units of $R_{25}$,
preventing to observe a drop in the mean corotation region (middle panel of Fig.~\ref{HistSDFSurfd}).
Therefore, most probably, the impact of DWs (triggering effect) is responsible
for a marginally higher surface density of CC~SNe within the mentioned radial range,
and for the inconsistency of the surface density distribution with
the inner-truncated exponential profile in LGD hosts
(middle panel of Fig.~\ref{HistSDFSurfd} and Table~\ref{tableGDNGDexp}).

To check the significance of the drop of surface density at $R_{\rm C}$
and excess at $\simeq 0.8$ and $1.5\,R_{\rm C}$ (see Fig.~\ref{surfdens}),
we study the distribution of CC~SNe distances to the nearest corotation
in units of corotation radius,
$D=|R_{\rm SN}-R_{\rm C}|/R_{\rm C}$.
Fig.~\ref{distCorot} displays the differential and cumulative distances
in the global disc of LGD galaxies.
Since a given value of distance can occur either for position $1-D$ or for position
$1+D$ (both in units of corotation radius),
the probability distribution function (PDF) of distances follows
\begin{eqnarray}
  {\rm PDF}(D) &=&
  \left\{
  \begin{array}{ll}
    \displaystyle
    f(1-D) + f(1+D)
    & 0 < D \leq 1 \\
    f(1+D) &
    D > 1
    \label{PDFD}
  \end{array}
  \right. \\
  f(x) &=& {x\over h^2}\,\exp(-{x\over h}) \ ,
\label{PDFD2}
\end{eqnarray}
where $h$ is the best-fitting scale length of the SNe (in
units of the corotation radii).
The CDF for $D$ is then
\begin{eqnarray}
  {\rm CDF}(D) &=&
  \left\{
  \begin{array}{ll}
    g(1-D) - g(1+D)
    & 0 < D \leq 1 \\
    1 - g(1+D)
    & D > 1
  \end{array}
  \right.
  \label{CDFD}
  \\
  g(x) &=& (1+ {x \over h})\,\exp(-{x \over h}) \ .
\label{CDFD2}
\end{eqnarray}

In Fig.~\ref{distCorot}, the black curves are the best-fitting (with MLE) expected distribution.
Fig.~\ref{distCorot} highlights the lack of CC~SNe at corotation and
excess outside/inside the $R_{\rm C}$ in the LGD hosts.
However, a KS test indicates a \emph{P}-value of
0.176, while an AD test indicates a \emph{P}-value of 0.197.
We check the significance of the drop/excess in the global disc,
adding also four CC~SNe from SGD sample.
The result is: $P_{\rm KS}=0.170$ and $P_{\rm AD}=0.224$.
For the inner-truncated disc of LGD (LGD+SGD) galaxies,
the $P_{\rm KS}=0.353 \, (0.445)$ and $P_{\rm AD}=0.299 \, (0.428)$.
Thus, the lack of CC~SNe at corotation and excess at $\simeq 0.8$ and $1.5\,R_{\rm C}$
do not appear statistically significant.
Note that these tests ignore the uncertainties
on the corotation radii. Including them would weaken even more the
statistical significance of these features in the surface density profile of CC~SNe.

It is important to note that, if one wants to test the star formation activity at the corotation,
the estimates of corotation radii based on kinematic or dynamic
arguments \citep[e.g.][]{2014ApJS..210....2F} would be preferable as they would be more independent of
the regions with lack of star formation \citep[e.g.][]{1992ApJS...79...37E}
or specific morphological features in the discs \citep[e.g.][]{2009ApJS..182..559B}.
Only 18 CC~SNe (17 in LGD and one in SGD) have host galaxies with such preferable estimates of $R_{\rm C}$.
If we consider only these objects, the triggering evidence and the dip in the global disc remain not significant
($P_{\rm KS}=0.514$ and $P_{\rm AD}=0.425$), probably due to even smaller statistics.

Another importance is that galaxies with several spiral patterns with different angular velocities,
i.e. more than one corotation, might have interactions between the patterns
\citep[e.g.][at $R_{\rm C}$ of one with inner/outer Lindblad resonance of the other]{2014ApJS..210....2F}
causing turbulence in the interface regions between the patterns and thereby increase
star formation activity at those regions \citep[see reviews by][]{2014PASA...31...35D,2016ARA&A..54..667S}.
Therefore, the distribution of CC~SNe (Figs.~\ref{surfdens} and \ref{distCorot})
might be contaminated by the objects at the $R_{\rm C}$, weakening the observed dip.
In Table~\ref{dataRcorGal}, we see that 31 CC~SNe (30 in LGD and one in SGD)
have host galaxies with single $R_{\rm C}$.
If we consider only these objects, the triggering evidence and the dip in the global disc
are again statistically not significant ($P_{\rm KS}=0.457$ and $P_{\rm AD}=0.354$).

Unfortunately, due to the insufficient number of CC~SNe in SGD galaxies
(only 4 objects, see Table~\ref{dataRcorGal}),
as well as Type Ia SNe in the LGD (4 cases) and SGD (4 objects) subsamples,
a similar study of their distributions relative to $R_{\rm C}$ is ineffective.
In the future, when more information is available on corotation radii of SN host galaxies,
we will be able to extend our study including all SN types in LGD and SGD galaxies.

\section{Conclusions}
\label{concl}

In this study, using a well-defined and homogeneous sample of SN host galaxies
from the coverage of SDSS DR13, we analyse the radial and surface density distributions
of Type Ia and CC~SNe in host galaxies with different ACs to
find the possible impact of spiral DWs as triggers for star formation.
Our sample consists of 269 relatively nearby (${\leq {\rm 150~Mpc}}$, the mean distance is 82~Mpc),
low-inclination ($i \leq 60^\circ$), morphologically non-disturbed and unbarred Sa--Sc galaxies,
hosting 333 SNe in total.
In addition, we perform an extensive literature search for corotation radii,
collecting data for 30 host galaxies with 56 SNe.

The main results concerning the deprojected and
inner-truncated ($\tilde{r} \geq 0.2$) distributions of SNe in host galactic discs
are the following:

\begin{enumerate}
\item We find no statistical differences between the pairs of
      the $R_{25}$-normalized radial distributions
      of Type Ia and CC~SNe in discs of host galaxies with different spiral ACs,
      with only one significant exception:
      CC~SNe in LGD and NGD galaxies have significantly different radius
      distributions (Table~\ref{RSNR25_KS_AD}).
      The radial distribution of CC~SNe in NGDs is
      concentrated to the centre of galaxies with relatively narrow peak and
      fast exponential decline at the outer region, while the distribution of
      CC~SNe in LGD galaxies has a broader peak,
      shifted to the outer region of the discs
      (upper panel of Fig.~\ref{HistSDFSurfd}).
\item The surface density distributions of Type Ia and CC~SNe in most of the subsamples
      are consistent with the exponential profiles.
      Only the distribution of CC~SNe in LGD galaxies appears to be inconsistent
      with an exponential profile (Table~\ref{tableGDNGDexp} for
      the AD statistic but only very marginally so for the KS statistic),
      being marginally higher at $0.4\lesssim R_{\rm SN}/R_{25} \lesssim0.7$.
      The inconsistency becomes more evident when comparing the same distribution
      with the scaled exponential profile of CC~SNe in NGD galaxies
      (middle panel of Fig.~\ref{HistSDFSurfd}).
\item Using a smaller sample of LGD galaxies with estimated corotation radii,
      we show, for the first time, that the surface density distribution of
      CC~SNe shows a dip at corotation, and enhancements at $^{+0.5}_{-0.2}$
      corotation radii around it (Fig.~\ref{surfdens}).
      However, these features are not statistically significant (Fig.~\ref{distCorot}).
      The CC~SNe enhancements around corotation may, if confirmed with larger
      samples, indicate that  massive star formation is triggered by the DWs in LGD host galaxies.
      Considering that the different LGD host galaxies have various corotation radii
      (Table~\ref{dataRcorGal} and Fig.~\ref{RcR25}) distributed around the mean value of
      $\langle R_{\rm C}/R_{25}\rangle=0.42\pm0.18$ (Fig.~\ref{RCR25histcum}),
      the radii of triggered star formation by DWs are most probably blurred
      within a radial region including $\sim0.4$ to $\sim0.7$ range in units of $R_{25}$,
      without a prominent drop in the mean corotation region (middle panel of Fig.~\ref{HistSDFSurfd}).
\end{enumerate}

These results for CC~SNe in LGD galaxies may, if confirmed with larger samples
and better corotation estimates, support the large-scale shock scenario
\citep[e.g.][]{1973PASP...85..564M}, originally proposed by \citet[][]{1969ApJ...158..123R},
which predicts a higher star formation efficiency, avoiding the corotation region
\citep[e.g.][]{1990ApJ...349..497C,2002MNRAS.337.1113S,2013AA...560A..59C,2016MNRAS.459.3130A}.

When more information will become available on corotation radii of SN host galaxies,
it would be worthwhile to extend our study, by comparing the $R_{\rm C}$-normalized
radial and surface density distributions of Type Ia and CC~SNe in LGD galaxies.
This will also allow to check the impact of spiral DWs
on the distribution of less-massive and longer-lived progenitors of Type Ia SNe.
Moreover, similar analysis of SNe in SGD galaxies can help to understand the role of DWs
in star formation triggering, if any.

\section*{Acknowledgements}

We would like to thank the referee, Preben Grosb{\o}l,
for excellent comments that improved the clarity of this paper.
AGK, AAH, and LVB acknowledge the hospitality of the
Institut d'Astrophysique de Paris (France) during their
stay as visiting scientists supported by
the Programme Visiteurs Ext\'{e}rieurs (PVE).
This work was supported by the RA MES State Committee of Science,
in the frames of the research project number 15T--1C129.
This work was made possible in part by a research grant from the
Armenian National Science and Education Fund (ANSEF)
based in New York, USA.
V.A. acknowledges the support from Funda\c{c}\~ao para
a Ci\^encia e Tecnologia (FCT) through Investigador FCT contract nr.
IF/00849/2015/CP1273/CT0003 and the support from FCT through national
funds and by FEDER through COMPETE2020 by the grant
PTDC/FIS-AST/7073/2014 \& POCI-01-0145-FEDER-016880.
Funding for the SDSS--IV has been provided by the Alfred P.~Sloan Foundation,
the US Department of Energy Office of Science, and the Participating Institutions.
SDSS--IV acknowledges support and resources from the Center for High-Performance Computing at
the University of Utah. The SDSS web site is \href{http://www.sdss.org/}{www.sdss.org}.
SDSS--IV is managed by the Astrophysical Research Consortium for the
Participating Institutions of the SDSS Collaboration including the
Brazilian Participation Group, the Carnegie Institution for Science,
Carnegie Mellon University, the Chilean Participation Group, the French Participation Group,
Harvard-Smithsonian Center for Astrophysics,
Instituto de Astrof\'isica de Canarias, The Johns Hopkins University,
Kavli Institute for the Physics and Mathematics of the Universe (IPMU) /
University of Tokyo, Lawrence Berkeley National Laboratory,
Leibniz Institut f\"ur Astrophysik Potsdam (AIP),
Max-Planck-Institut f\"ur Astronomie (MPIA Heidelberg),
Max-Planck-Institut f\"ur Astrophysik (MPA Garching),
Max-Planck-Institut f\"ur Extraterrestrische Physik (MPE),
National Astronomical Observatories of China, New Mexico State University,
New York University, University of Notre Dame,
Observat\'ario Nacional / MCTI, The Ohio State University,
Pennsylvania State University, Shanghai Astronomical Observatory,
United Kingdom Participation Group,
Universidad Nacional Aut\'onoma de M\'exico, University of Arizona,
University of Colorado Boulder, University of Oxford, University of Portsmouth,
University of Utah, University of Virginia, University of Washington, University of Wisconsin,
Vanderbilt University, and Yale University.

\bibliography{snbibDW}

\section*{Supporting information}

Supplementary data are available at \emph{MNRAS} online\\
\\
\textbf{Onlinedata.csv}
\\
\\
Please note: Oxford University Press is not responsible for the
content or functionality of any supporting materials supplied by
the authors. Any queries (other than missing material) should be
directed to the corresponding author for the article.

\label{lastpage}

\end{document}